\documentclass[fleqn,usenatbib,useAMS]{mnras}

\usepackage{newtxtext,newtxmath}

\usepackage[T1]{fontenc}

\usepackage{graphicx}	
\usepackage{amsmath}	
\usepackage{multicol}        
\usepackage{bm}		
\usepackage{longtable}
\usepackage{xtab}
\usepackage{rotating}

\newcommand{\gaia}{\emph{Gaia}}

\newcommand{\teff}{\ensuremath{T_{\rm eff}}}

\newcommand{\bprp}{$B_{\rm P}-R_{\rm P}$} 

\usepackage[T1]{fontenc}
\usepackage{ae,aecompl}
\usepackage{breqn}

\usepackage{newtxtext,newtxmath}
\usepackage{pdflscape}	

\title[Age of the Beta Pictoris Moving Group]{Revisiting the Membership, Multiplicity, and Age of the Beta Pictoris Moving Group in the \gaia\ Era}

\author[R. A. Lee et al.]{Rena A. Lee\thanks{Contact e-mail:renaalee@hawaii.edu}$^{1,2}$,
Eric Gaidos$^{1,3,4}$,
Jennifer van Saders$^{2}$,
Gregory A. Feiden$^{5}$
\newauthor
and Jonathan Gagn\'{e}$^{6,7}$
\\
$^{1}$Department of Earth Sciences, University of Hawai'i at M\={a}noa, Honolulu, HI  96822, USA\\
$^{2}$Institute for Astronomy, University of Hawai'i at M\={a}noa, Honolulu, HI 96822 USA\\
$^{3}$Institute for Astrophysics, University of Vienna, 1180 Wien, Austria\\
$^{4}$Institute for Particle Physics \& Astrophysics, ETH Z\"{u}rich, 8093 Z\"{u}rich, Switzerland\\
$^{5}$Department of Physics and Astronomy, University of North Georgia, Dahlonega, GA 30597 USA\\
$^{6}$Planétarium Rio Tinto Alcan, Espace pour la Vie, Montr\'{e}al, Qu\'{e}bec, Canada\\
$^{7}$Institute for Research on Exoplanets, Universit\'{e} de Montr\'{e}al, Montréal, Qu\'{e}bec, Canada
}
\date{Submitted, accepted}

\pubyear{2022}

\begin{document}
\label{firstpage}
\pagerange{\pageref{firstpage}--\pageref{lastpage}}
\maketitle

\begin{abstract} 
Determining the precise ages of young (tens to a few hundred Myr) kinematic (``moving") groups is important for placing star, protoplanetary disk, and planet observations on an evolutionary timeline. The nearby $\sim$25 Myr-old $\beta$ Pictoris Moving Group (BPMG) is an important benchmark for studying stars and planetary systems at the end of the primordial disk phase. \gaia\ DR3 astrometry and photometry, combined with ground-based observations and more sophisticated stellar models, permit a systematic re-evaluation of BPMG membership and age.  We combined \gaia\ astrometry with previously published radial velocities to evaluate moving group membership in a Bayesian framework.  To minimize the effect of unresolved stellar multiplicity on age estimates, we identified and excluded multi-star systems using \gaia\ astrometry, ground-based adaptive optics imaging, and multi-epoch radial velocities, as well as literature identifications. We estimated age using isochrone and lithium-depletion-boundary fitting with models that account for the effect of magnetic activity and spots on young, rapidly rotating stars. We find that age estimates are highly model-dependent; Dartmouth magnetic models with ages of 23$\pm$8 Myr and 33$^{+9}_{-11}$ Myr provide best fits to the lithium depletion boundary and \gaia\ $M_G$ vs. $B_{P}$-$R_{P}$ color-magnitude diagram, respectively, whereas a Dartmouth standard model with an age of 11$^{+4}_{-3}$ Myr provides a best fit to the 2MASS-\gaia\ $M_{K_S}$ vs. $B_{P}$-$R_{P}$ color-magnitude diagram.
\end{abstract}


\section{Introduction}
\label{sec:intro}

Associations of nearby, kinematically coherent, coeval stars, known as young moving groups (YMGs), are indispensable laboratories for studying early stellar and planetary evolution. Studies of YMG members have informed research on early evolution of stellar interiors, rotation, magnetic activity, and high-energy emission \citep{Kastner2022}, protoplanetary disk-dissipation timescales \citep{Silverberg2020,Higashio2022}, and the occurrence of giant planets \citep{Nielsen2019}.  Accurately describing these evolutionary sequences relies on precise and accurate ages for these groups, i.e., by fitting model isochrones to stars in a color-magnitude diagram,  modeling the depletion of lithium in the lowest-mass members, or back-tracking stars to a common spatial origin based on their kinematics.  The ages of individual members (and their planets) can then be assigned to the ensemble age of the host YMG\footnote{This assumes star formation occurred over an interval that is short compared to the age of the group.}.  In recent years, the advent of all-sky photometric and astrometric surveys, especially by the \gaia\ mission, has enabled the identification of new members of known YMGs, as well as discovery of new groups \citep{Faherty2018,Gagne2018b,Ujjwal2020,Lee2022}.  

The $\beta$ Pictoris Moving Group (BPMG), as the youngest, nearest (median distance $\sim$50 pc) YMG, permits some of the most detailed studies of pre-main sequence stars (pre-MS), including the intrinsically faintest (lowest and substellar-mass) members.  Projected angular separations between binaries are larger, facilitating the identification and characterization of stellar and substellar companions to the members. The age of the group, estimated to be $\sim$25 Myr by lithium depletion boundary (LDB) fitting \citep{Messina2016}, isochrone fitting, and dynamical traceback \citep{Mamajek2014}, corresponds to an epoch immediately following the dissipation of protoplanetary disks and the initial phase in the evolution of fully-formed stars and planets.  The BPMG's eponymous star, $\beta$ Pictoris, hosts one of the first directly imaged debris disks \citep{Lecavelier1993}, as well as a rare directly imaged multi-planet system \citep{Lagrange2020}. Among other BPMG members are 51 Eridani, hosting a directly imaged Jupiter analog \citep{Macintosh2015}, PSO~J318.5-22, a free-floating planet \citep{Liu2013}, and AU Microscopii, hosting an edge-on debris disk and two transiting planets \citep{Kalas2004,Plavchan2020,Cale2021}. 

\begin{figure}
    \centering
    \includegraphics[width=\columnwidth]{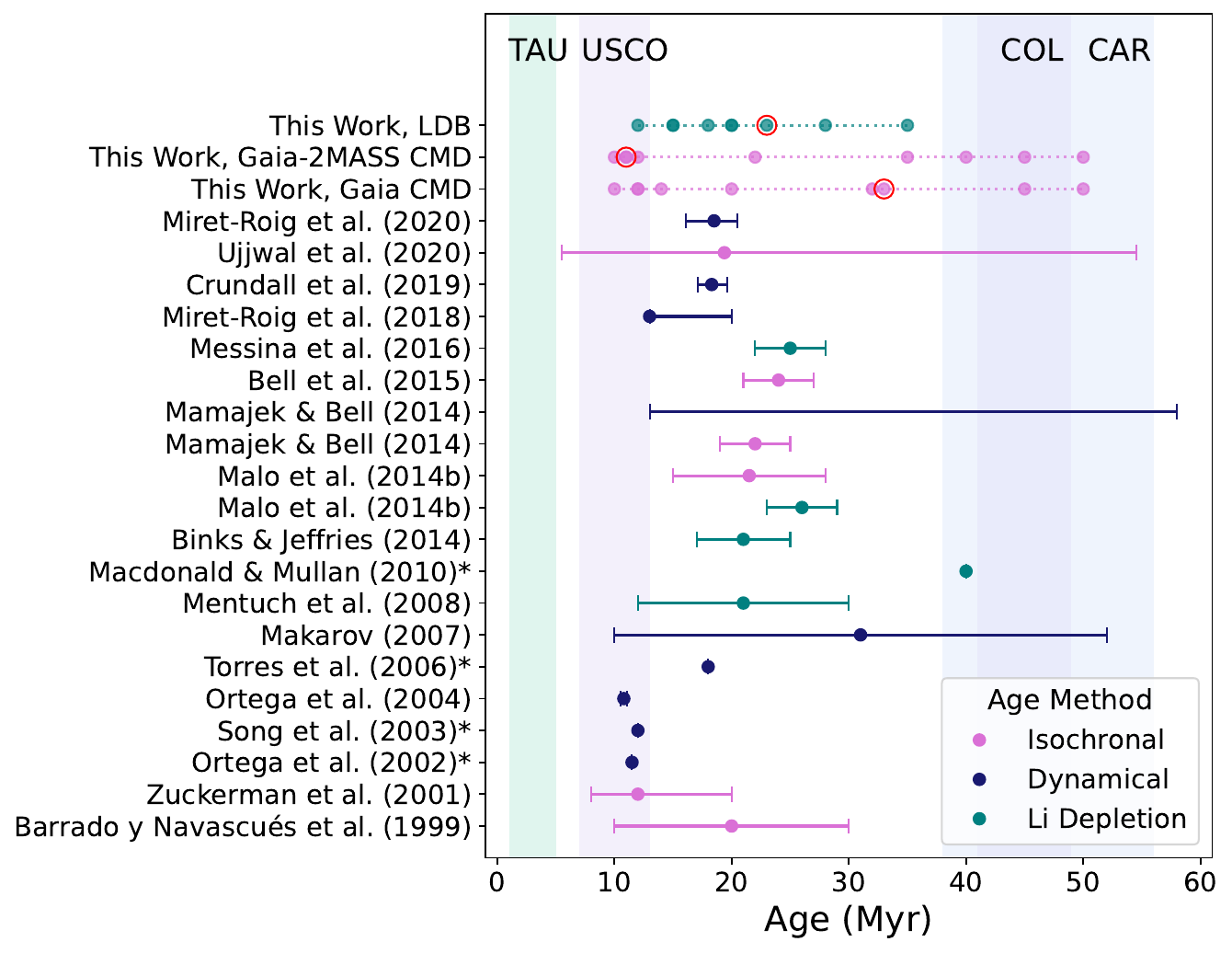}
    \caption{Estimates of the age of the BPMG in the literature, i.e. Table 6 of \citet{Miret-Roig2020}, updated from Table 1 of \citet{Mamajek2014}, along with the CMD- and LDB-based ages from this study. The best-fit ages for each of three methods used in this work are circled in red. Estimated ages of other young stellar associations Taurus (TAU), Upper Scorpius (USCO), Carina (CAR), and Columba (COL) from Table 1 in \citet{Gagne2018b} have been plotted for comparison and context.   The sources for ages marked with a "*" do not report errors for their ages.}
    \label{fig:bpmg_ages}
\end{figure}

The age of the BPMG has been estimated by fitting model isochrones to sequences in color-magnitude diagrams (CMDs), dynamical traceback analyses, and lithium-depletion-boundary (LDB) model fitting.  The derived values range between 10 and 50 Myr, with considerable variation in the uncertainties \citep[Fig. \ref{fig:bpmg_ages},][]{Mamajek2014}. The most recent estimates cluster around 20-25 Myr, placing the BPMG at an intermediate stage between the Upper Scorpius star-forming region \citep[10$\pm$3 Myr,][]{Pecaut2016}, and the Carina YMG at 38-56 Myr \citep{Bell2015,Gaidos2022c}. Dynamical traceback studies of BPMG kinematics suggest a single star-forming origin \citep[e.g.,][]{Crundall2019,Miret-Roig2020}, and thus is no evidence for a physical age spread among BPMG members. 
 
Several effects can contribute to systematic error in the estimated ages of moving-groups.  Contamination by non-members from the field can increase the scatter in color-magnitude diagrams and cause a group to appear older. Stars in other, older moving groups within the solar neighborhood, which are more likely to have space velocities more similar to BPMG, may also erroneously be identified as members.  Non-members contaminating the sample can be identified and removed based on Galactic space motions, but this requires precise parallaxes, proper motions, and radial velocities (RVs).   Unresolved binary companions can make members appear more luminous and hence younger than they actually are on CMDs \citep{Sullivan2021}, affect the kinematics used for trace-back analysis, or influence the depletion of Li through the effect of a stellar companion on disk dissipation and stellar rotation \citep{Somers2015}. Previous studies have highlighted the importance of identifying and excluding binary systems from group age analyses \citep[e.g.,][]{Binks2014,Alonso-Floriano2015,Messina2017}.  Recently, \citet{Miret-Roig2020} used \gaia\ DR2 astrometry and ground-based RVs to assess the BPMG age, identifying binaries based on the astrometry and RV variability.  

An additional source of uncertainty in deriving ages from model comparisons comes from the elevated activity of members of young groups. At young ($<$ 0.5 Gyr) ages, most low-mass ($<$ 0.8 M$_\odot$) stars are rapidly rotating (P $<$ 10 d), and the surface spottedness and magnetic activity of pre-MS stars are observed to be enhanced compared to more quiescent MS stars \citep{Morris2020}.  It is not known if any observed internal age gradients within stellar groups are physical, as the mass-dependence of star formation timescales is not well established. Multiple epochs of star formation may contribute to actual scatter in group age estimates \citep{Krumholz2019}, but as mentioned, this case has not been supported by dynamical studies of BPMG \citep{Crundall2019,Miret-Roig2020}.

An individual star's membership in a YMG is typically assessed by comparing its position in Galactic 6-d (3-d position and 3-d space velocity) phase space to that of the established group mean and dispersion. Previous BPMG membership studies have cataloged candidate members \citep[e.g.,][]{Gagne2018b} using incomplete astrometric (5-D; 3-D position and 2-D motion) information, followed by measurement of the 6th (RV) dimension to confirm membership \citep[e.g.,][]{Schneider2019}.  Additionally, several studies have used RVs to identify binaries among BPMG members \citep{Alonso-Floriano2015,Messina2017,Miret-Roig2020}. However, multi-epoch observations were often not available, meaning that single-lined spectroscopic binaries with variable RVs could not be identified.  

The advent of the \gaia\ mission \citep{Gaia2016} has provided the precise astrometry and, for many sufficiently bright stars, RVs, to calculate space motions and vet members of YMG.  Incomplete RV information can be complemented by ground-based surveys such as the Radial Velocity Experiment \citep[RAVE,][]{Steinmetz2020}. \gaia\ also provides precise and homogeneous optical photometry, i.e. magnitudes in the $G$-band (similar to but much wider than Cousins $R$) and a \bprp\ color constructed from its wavelength-dispersing photometer.   These can be used to construct color-magnitude diagrams to compare with model isochrones, and to identify unresolved binaries or contaminated single-star sources via over-luminosity relative to known single stars.  \gaia\ Data Release 3 provides a metric of astrometric error, the Reduced Unit Weight Error (RUWE), relative to a single-star solution, that can be used to identify unresolved binaries \citep{Belokurov2020,Wood2021}.  This can be complemented by high-angular resolution adaptive optics (AO) imaging obtained by searches for substellar companions; these can readily detect stellar companions at  separations of $\sim$10-1000 au.

New, diverse evolutionary models are available to compare with photometry and measurements of Li abundance, in particular those addressing the effects of magnetic activity and starspots on the structure, convection, and luminosity of low-mass pre-MS stars \citep{Feiden2016,Somers2020}.   The contribution of magnetic pressure to the internal gas equation of state and total near-surface pressure leads to inflated predicted radii for a given mass and age relative to non-magnetic models, or, for a given luminosity and effective temperature, to an older model age.     

In this work, we combined precision astrometry, photometry, and RVs from \gaia\ DR3 with ground-based RV measurements and AO imaging to assess and confirm moving group membership, identify and exclude close binaries, and distill a catalog of single-star members of the BPMG suitable for age analysis by the CMD and LDB methods. Further, we compared our estimate of the multiplicity fraction from this work with those of other young stellar associations. We fit predictions from a suite of  evolutionary models designed for spotted, magnetically active stars to precise, homogeneous measurements of these stars to revisit the age of the BPMG, and consider the implications of our new, model-dependent ages. 

\section{Observations and Methods}
\label{sec:observations}

\subsection{Catalog Construction}
\label{sec:catalog}

We obtained confirmed and candidate members of the BPMG from the \citet{Gagne2018a} catalog, which analyzed all \gaia\ DR2 sources with precise parallaxes within the nearest 100 pc for YMG membership. We screened additional confirmed and candidate members from \citet[]{Binks2014}, \citet{Binks2016}, \citet{Kiss2011}, \citet{Lee2018}, \citet{Messina2017}, and \citet{Shkolnik2017}, yielding an initial catalog of 433 (assumed individual) stars.  Binaries are discussed in  Sec. \ref{sec:binaries}.  This sample was cross-matched with \gaia\ DR3 \citep{Gaia2021} and astrometry and photometry were incorporated for 415 stars and RVs for 99 stars. In total, RVs of 238 candidates were gathered, from RAVE DR6 \citep{Guiglion2020}, and previous RV surveys of low-mass stars and YMGs \citep[i.e.,][]{Binks2016,Malo2014a,Shkolnik2017,Terrien2015}, in addition to the \gaia\ RVs.  

 We calculated probabilities of BPMG membership using {\tt BANYAN} $\Sigma$, a multivariate Bayesian classifier \citep{Gagne2018a}.  {\tt BANYAN} $\Sigma$ calculates $UVW$ space velocities and $XYZ$ galactic positions from coordinates, proper motions, parallaxes, RVs, and their errors, and compares these to the six-dimensional group means and standard deviations of the 27 nearest identified young moving groups. For candidates with multi-epoch RV measurements, we used the median values in the {\tt BANYAN} inputs. For candidates with an even number of epochs, the mean of the median values were used. Based on the distribution of membership probabilities computed for candidates with complete kinematic data, we chose a threshold probability of 0.9; this selected 148 of 238 candidates with RV measurements.  Separately, we identified 63 of the 195 candidates lacking RVs which would have membership probabilities $>$0.9 for an ``optimal" RV that maximizes that probability and is plausible, i.e., is within the distribution of the measured RVs of confirmed members.  These are high-priority targets for ongoing RV follow-up but are not further analyzed in this work.

\begin{figure*}
    \centering
    \includegraphics[width=\textwidth]{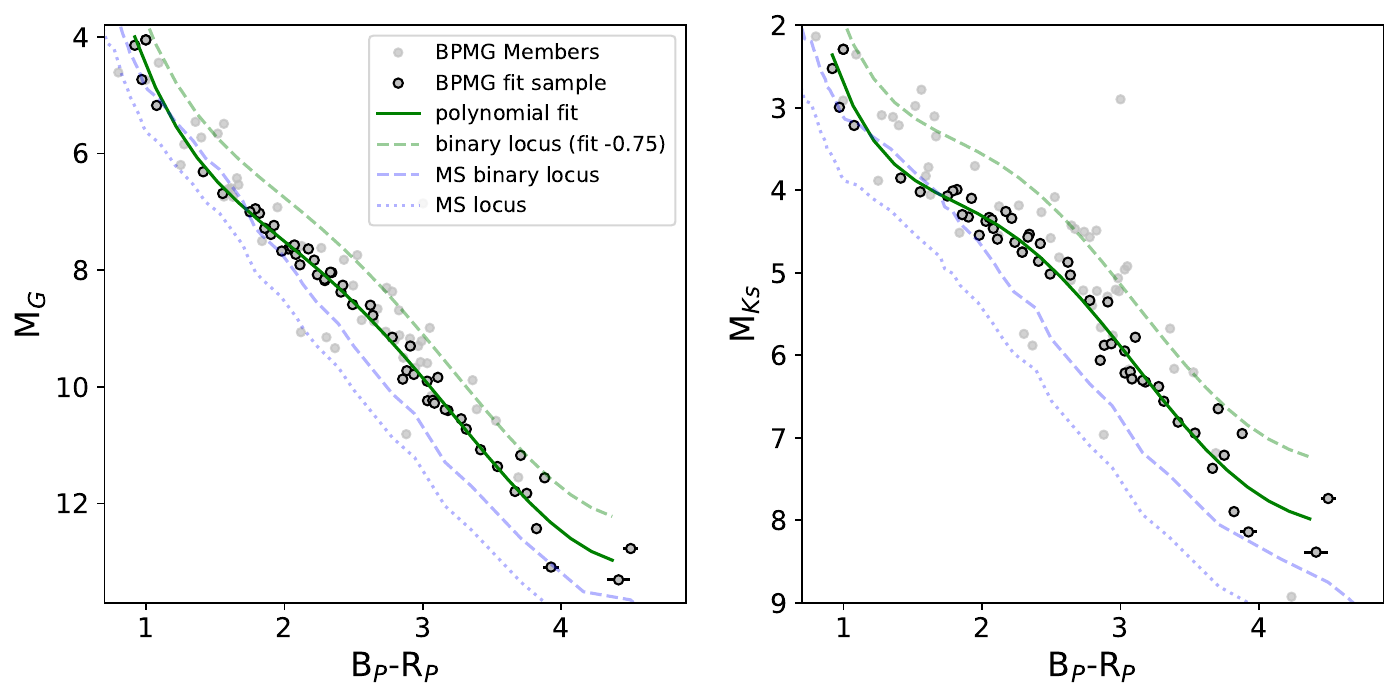}
    \caption{\textit{left:} Color-magnitude diagram (CMD) based on \gaia\ absolute $G$ mangitudes and \bprp\ colors of confirmed BPMG members.  \textit{right:} CMD using absolute 2MASS $K_s$ magnitude in place of $M_G$.   All confirmed BPMG members are shown as grey dots, while those passing the binary exclusion and photometric criteria and used for age estimation (outlined in black) are used in the isochrone fitting described in Sec. \ref{sec:cmd}.   The solid green lines are the best-fit lines in each diagram (5th degree polynomial) and the dashed green lines are the equal-mass binary loci (displaced by $-2.5 \log 2 \approx -0.75$ mag from the best-fit loci).  The blue dashed line is the equal-mass binary locus constructed from the empirical main sequence (blue dotted line) of \citet{Pecaut2013}. }
    \label{fig:cmds}
\end{figure*}
 
Main sequence stars could be interlopers if their space motions coincide by chance with the BPMG.  To identify and filter these, we compared $p>0.9$ candidates with the empirical main sequence of \citet{Pecaut2013} in a absolute $G$ magnitude vs. \bprp\ color CMD (Fig. \ref{fig:cmds}).  We computed the minimum offset in $G$ magnitude of each candidate from the MS. Based on the distribution of offsets, we chose a cutoff of 0.55 mags and removed nine stars within 0.55 mags of the MS locus and \bprp $>2$, corresponding to a mass of $<0.55 M_{\odot}$ \citep{Pecaut2013}, as interlopers. We retained stars with \bprp $<2$ since hotter, more massive stars can be close to or on the MS by the age of the BPMG \citep{Krumholz2019}. 

We cross-matched BPMG stars with the 2-Micron All-Sky Survey (2MASS) Point Source Catalog \citep{Skrutskie2006} to obtain $JHK_s$-band photometry in the infrared, and we corrected for extinction and reddening based on the \citet{Leike2020} 3-d ISM dust distribution using {\tt Dustmaps} \citep{Green2018}.  As expected for nearby stars, ISM reddening is small; the mean reddening is $E(B-V) =0.0005$ mag, and the mean extinction values are $A_G =  0.0003$ mag and $A_{K_s} = 7.1\times10^{-5}$ mag.

The 4" resolution of the 2MASS survey means that many binary stars will be unresolved.  Unresolved or partially resolved sources can lead to erroneously brighter and somewhat redder photometry that in turn can affect age-dating \citep{Sullivan2021}.  To minimize this systematic, we corrected the $K_s$-band photometry of those binaries that were resolved by \gaia\ (see Section \ref{sec:binaries}) but not resolved by 2MASS (noted in Table \ref{tab:members}).  We predicted the 2MASS $K_s$ magnitude of each \gaia-resolved component using the $G$ magnitude of each compnent and a relation between $G-K_s$ and \bprp\ based on the empirical main sequence of \citet{Mamajek2014}\footnote{This color-color relation is based on main sequence stars and ignores metallicity effects but should be approximately correct for BPMG members.}.  We then compared the predicted and the measured $K_s$ mags, and based on the overall distribution of the differences, we flagged systems with a difference $>$0.25 mag as significantly contaminated by a companion. Based on the contrast of the predicted $K_s$ magnitudes, we apportioned the measured $K_s$-band fluxes between the binary components, and apparent $K_s$ mags from those fluxes. This salvaged 9 systems for use in the $K_s$ vs. \bprp\ CMD fitting in Sec. \ref{sec:cmd}.

\subsection{Binary Identification}
\label{sec:binaries}

\begin{figure*}
    \centering
    \includegraphics[width=\textwidth]{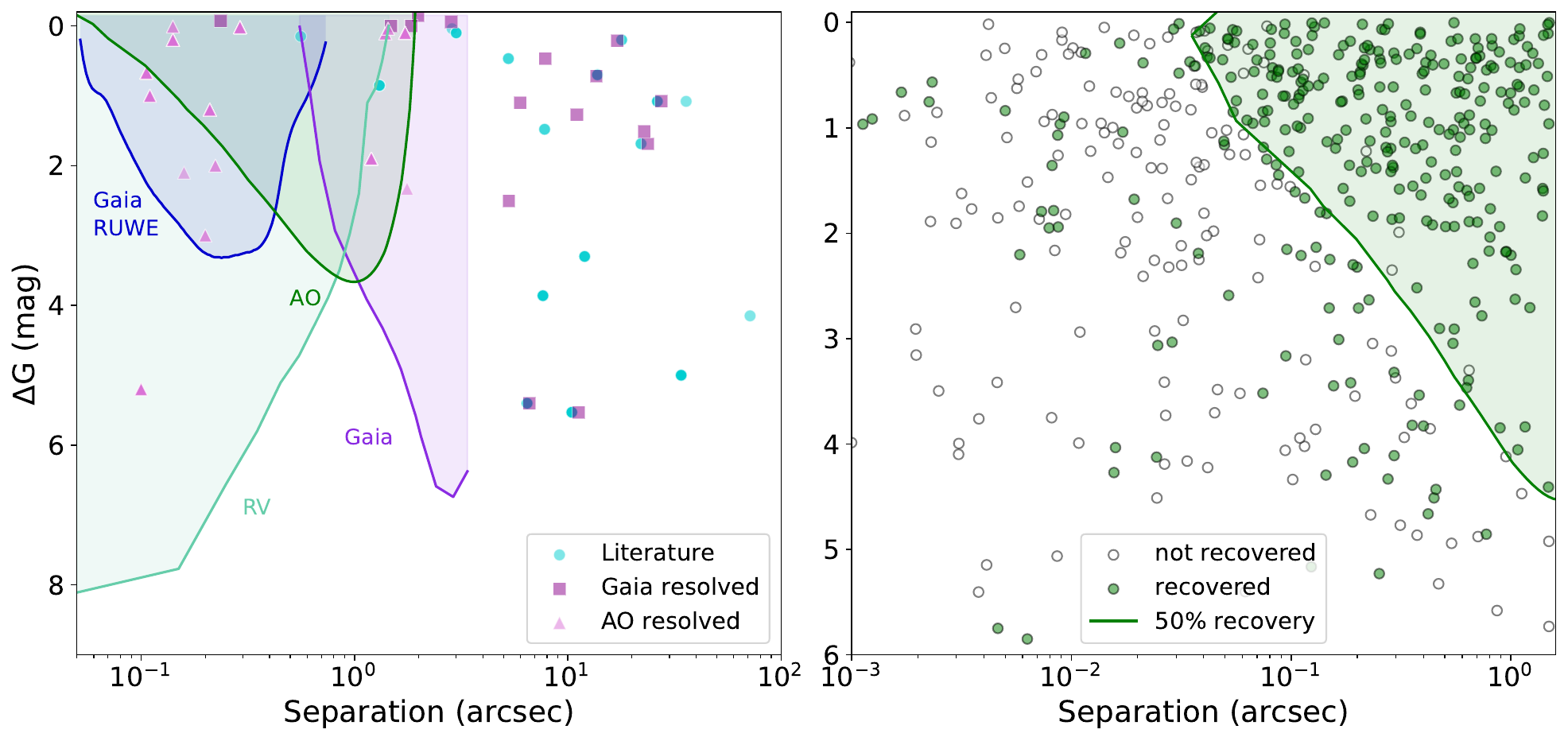}
\caption[Identified Resolved Binaries]{\textit{Left}: Separation vs. magnitude contrast for binary components in the BPMG primary sample from the present \gaia\ cone search astrometry comparison and AO image inspection, as well as literature-reported binaries. Unresolved photometric, spectroscopic, and RUWE-identified binaries are not included since these parameters cannot be unambiguously determined.  Also plotted are 50\% recovery contours for \gaia\ RUWE binaries from \citet{Wood2021}, \gaia\ -resolved binaries from \citet{Ziegler2018} and for AO from our injection-and-recovery tests. The estimated RV detection limit, based on an assumed primary mass of 1 $M_\odot$ and average RV amplitude of 20 km s$^{-1}$, is shown in light green. \textit{Right}: Results of the AO injection-and-recovery analysis. Simulated binaries that were not recovered by eye (false negatives) are plotted as open points, and successful recoveries are plotted as filled points. The contour represents a 50\% detection efficiency.}
\label{fig:coverage}
\end{figure*}

We retrieved visual and spectroscopic binary systems from \citet{Shkolnik2017} and spectroscopic and RV binary systems from \citet{Miret-Roig2020}.  We identified additional binary/multiple systems that appear as \gaia-resolved components in the DR3 astrometric catalog, unresolved systems identified by \gaia\ astrometric error, systems that are resolved in high-resolution AO imaging, and unresolved systems via multi-epoch RV variability or over-luminosity relative to the single-star locus in a CMD.  Fig. \ref{fig:coverage} shows the approximate coverage of these different methods in magnitude contrast-separation space. 

 
\textit{Gaia resolved binaries:} We identified sources within 30" with similar parallaxes and proper motions; this search radius corresponds to a binary separation of 1500~au at the median distance of BPMG members, well beyond the peak of the M dwarf binary separation distribution at 50~au \citep{Susemiehl2022}, or $\sim$5~au \citep{Winters2019}, and thus encompassing the vast majority of physical companions.  For each pair of member and potential companion we computed likelihoods $P_1$ and $P_2$ that the latter is a common proper-motion companion or an interloper from the field, respectively. We then compute a Bayesian probability $P_c/(P_c+P_f)$, where $P_{c,f}$ are both evaluated as:
\begin{equation}
\label{eqn:prob}
P = \sum_{i} \Delta \mu_i \times \exp\bigg(-\frac{\Delta\mu_i^2}{2} \frac{1}{\sigma_\mu^2 + \mu_{\rm orb}^2}  \\
 -\;\frac{\Delta \pi^2}{2\sigma_\pi^2}\bigg)\; \frac{1}{\sigma_\mu^2+\mu_{\rm orb}^2} \frac{1}{\sigma_\pi}, 
\end{equation}

\noindent except that while the sum for $P_c$ consists only of a contribution from  a single putative companion, the sum for $P_f$ is over all stars from an equivalent search at an uncorrelated (distant) location from the BPMG member.  In Eqn. \ref{eqn:prob}, $\Delta \pi_i$ and $\Delta \mu_i$ are the difference in parallax and proper motion, respectively, $\sigma_\pi$ and $\sigma_\mu$ are the standard errors in parallax and proper motion of a stellar pair, added in quadrature, and $\mu_{\rm orb}$ is the typical proper motion due to orbital motion, included as a ``softening" term. This was estimated by considering the peak separation of nearby ($<$ 25 pc) M dwarf binaries, 5~au \citep{Winters2019}, and 2$\pi \cdot 5 \approx 30$~au orbit. For missions such as \gaia, a five-year baseline would allow detection of $\sim$6 au yr$^{-1}$ orbital motion, which at the BPMG's median distance of 50 pc, is on the order of 100 mas yr$^{-1}$. Because of orbital inclinations and eccentricities, realistically there would be a decrement in the observed orbital motions on the order of a half. For our computations, we adopted $\mu_{\rm orb}=10$ mas yr$^{-1}$, which follows from an orbital motion of $\sim$0.5 au yr$^{-1}$. Wider separations would correspond to smaller orbital motions that are not detectable. 

Based on the distribution of $P$ values we adopted a cutoff of $P > 0.98$.  Our Bayesian analysis identified 30 likely multi-star systems; 29 binaries and one triple. We confirmed 14 of these companions as BPMG members based on their astrometry, independent RVs, and {\tt BANYAN} $\Sigma$. The separations and magnitude contrasts of these 14 systems are shown in Fig. \ref{fig:coverage} as purple squares, labeled "\gaia\ Resolved."  Binary systems resolved by \gaia\ and 2MASS were included in subsequent analyses; those resolved by \gaia\ but not by 2MASS were included if the latter were resolved with AO imaging (see below) or corrected for source confusion as described above.
 
\textit{AO imaging:} We collected archival AO images of confirmed BPMG members obtained with any of three instruments: NAOS+CONICA (NACO) on the ESO VLT UT1 \citep[][]{Lenzen2003,Rousset2003}, NIRC2 on Keck-2 \citep[]{Service2016}, and NIRI on Gemini-North \citep{Hodapp2003} from the ESO, Keck Observatory, and Gemini Observatory archives, respectively. In each archive, we performed a 10" cone search (the typical field of view) of the coordinates of each member. We retrieved at least one AO image for 117 BPMG members, and we inspected each image for additional sources.  In Fig. \ref{fig:coverage} (left) we plot the 16 AO-resolved binary components as pink triangles.  An injection-and-recovery analysis was performed to assess the completeness and false positive rate of our visual inspection, i.e. by adding simulated companions to randomly selected AO images containing a single (real) star.  The point spread function of the target star was scaled and used for the simulated (mock) companion. The magnitude contrast of the simulated companion was randomly sampled from a uniform distribution in magnitude, and the separation was sampled from the \citet{Winters2019} orbital distribution of nearby M dwarf systems, assuming a distance of 50 pc. Separations were limited to 1.5" to focus on a region of the parameter space not resolved by \gaia; the recovery and false-positive rates at wider separations is expected to be the same as at 1.5" out to at least $\sim$5", half the field of view of the NIRC2 images.  The results of the data injection recovery are shown in Fig. \ref{fig:coverage} (right).  We performed a kernel density estimation using the {\tt Python} package {\tt sklearn.neighbors.KernelDensity} \citep{scikit-learn} on the recovered (green points) and not-recovered (open points) simulated companions to establish a 50\% detection efficiency contour in separation-magnitude contrast space (Fig. \ref{fig:coverage} (right)).  From the injection-and-recovery analysis, we found the visual inspection was reliable for detecting companions having up to 4.5 magnitudes of contrast with the primary and separation as small as 40 mas.

 \textit{Gaia astrometric error:}  \gaia\ RUWE is a measure of astrometric error relative to a single-star fit that is analogous to reduced $\chi^2$. Stars with RUWE $>$ 1.4 are almost invariably unresolved binaries \citep[A. Kraus, pers. comm.,][]{Belokurov2020}.  \citet{Fitton2022} found that single stars hosting protoplanetary disks also exhibit elevated RUWE, but BPMG members are known to host only much less substantial debris disks \citep{Lagrange2010}.  Twenty-nine members were identified as \gaia-unresolved binaries based on RUWE.
 
\textit{Radial velocities}: Sources with ground-based RVs at more than one epoch were assessed for binarity based on a reduced $\chi^2$ test.  Based on the distribution of values, we chose a criterion of $\chi^2_{\nu} > 20$, flagging 6 systems. We show in Fig. \ref{fig:coverage} an estimated RV detection limit contour based on the separations and magnitude contrast of the resolved BPMG binaries and the average RV amplitude from members with multi-epoch RVs. We computed the expected separation and magnitude contrast of unresolved binary systems assuming a typical primary mass of 0.3 M$_\odot$ (for M dwarfs) and maximum separation of 1000 au. Additionally, the probability of the null hypothesis ($p$-value) for RV variability are included for many sufficiently bright stars in \gaia\ DR3 \citep{Chance2022} and 14 sources for which $p< 0.01$ were flagged as spectroscopic binaries. 

\textit{Photometry}: Additional candidate binaries were identified by the over-luminosity of stars with respect to the nominal single-star locus for BPMG (described in Sec. \ref{sec:catalog}).  We assumed the stars to be equal-brightness, the most probable configuration among M dwarfs \citep{Duchene2013} and constructed a binary locus for the BPMG by shifting the single-star locus by -0.75 mag. We quantified the absolute offset in $M_G$, $\Delta M_G$, of individuals from this locus and identified 18 likely photometric binaries based on a cut-off of $\Delta M_G\;<$ 0.7 mag. 

In summary, our membership catalog began with 433 candidate sources compiled from the literature, of which 238 were tested for BPMG membership with full kinematic data. Among the 148 confirmed members, we identified 30 resolved binary systems; 14 from the \gaia\ proper motion matching and 16 from AO imaging. We identified and excluded 61 unresolved binary systems from subsequent analyses; 29 from \gaia\ RUWE, 14 from RVs, and 18 from photometry. Since they are not resolved, they are not shown in Fig. \ref{fig:coverage} (left), though we provide approximate detection limits for these methods. This left 64 single stars and 30 resolved binary systems for subsequent analyses.

\section{Age Determination}
\label{sec:ages}

\subsection{Isochrone Fitting}
\label{sec:cmd}

We estimated the age of the BPMG by fitting isochrones from different model suites to the confirmed members that are single or resolved binaries in the two CMDs plotted in Fig. \ref{fig:cmds}. We assessed four combinations of \gaia\ and 2MASS colors and magnitudes: $M_G$ vs. \bprp, $M_{K_s}$ vs. \bprp, $M_G$ vs. G-K, and $M_{K_s}$ vs. G-K. To quantify the scatter in each locus, we computed the $\chi^2$ offsets from best-fit 5th order polynomials for each CMD. The final sample for isochrone fitting is plotted in $M_G$ vs. \bprp\ and $M_{K_s}$ vs. \bprp\ CMDs (Fig. \ref{fig:cmds}), which are the two best-performing choices of color-magnitude pairs for CMD analysis.  We used the pre-MS models of \citet{Baraffe2015}, the Dartmouth standard and magnetic models \citep{Feiden2016}, and the SPOTS models of \citet{Somers2020}.  The \citet{Baraffe2015} models (hereafter, BHAC) are calibrated for pre-MS and MS low-mass stars. The evolutionary calculations are based on standard input physics of stellar interiors \citep{Chabrier1997}, but with updated atmosphere models compared to the predecessor evolutionary models of \citet{Baraffe1998}. While the BHAC models are calibrated for pre-MS stars, they do not account for the effects of stronger magnetic fields occurring at the surfaces of young, rapidly rotating stars. We interpolated model values of \gaia\ and 2MASS magnitudes over the provided masses and ages to produce a finer age model grid (spaced by 1 Myr) than is provided in \citet{Baraffe2015} to better intercompare the precise colors and absolute magnitudes.  

Dartmouth stellar evolution models are suitable for modeling low-mass main sequence \citep{Dotter2008} and pre-MS stars \citep{Feiden2016}, with or without the effects of stellar magnetic fields \citep{Feiden2012,Feiden2016}. The magnetic and non-magnetic models were used to age-date BPMG in \citet{Malo2014b}. Their interior structure and evolution calculations adopt standard input physics (e.g., solar-calibrated mixing length, general ideal gas equation-of-state, and MARCS \citep{Gustafsson2008} model atmospheres) comparable to other modern low-mass stellar models \citep[e.g., BHAC;][]{Dotter2008}, with updates to improve the accuracy of model predictions for very-low-mass stars and the pre-main-sequence phase of stellar evolution \citep[e.g., ][]{Malo2014b,Feiden2015}. Models including magnetic fields account for the impact of magnetism on the gas equation of state and on the efficiency of thermal convection \citep{Feiden2012,Feiden2016}. Surface magnetic field strengths are assumed to be in pressure equipartition with the gas near the stellar photosphere \citep{Feiden2016}. The model grid assumes solar metallicity (appropriate for the BPMG) and solar composition \citep{Grevesse2007} and includes individual mass tracks with $0.1 \le M/M_\odot \le 2.0$, and isochrones with ages from 1 Myr -- 10 Gyr. Synthetic \gaia\ photometry is calculated using synthetic spectra from MARCS model atmospheres \citep{Gustafsson2008}, 2MASS $JHK_s$ passbands \citep{Skrutskie2006}, and revised \gaia\ DR2 passbands \citep{Evans2018} assuming the extinction $A_V$ is zero. 

SPOTS evolutionary models \citep{Somers2020} address the distorting effects of activity and surface spots on the structure of magnetically active stars. The model grid includes evolution tracks and isochrones for 0.1 -- 1.3 M$_\odot$ stars with discrete surface spot coverage fractions of \textit{f} = 0.0, 0.17, 0.34, 0.51, 0.68, and 0.85. The SPOTS isochrones are empirically calibrated from the color transformations of \citet{Pecaut2013}. However, \gaia\ magnitudes and colors are not provided for objects redder than \bprp\ $\sim$2.5 for models and $f>$ 0, where many BPMG members lie.  We extend the fits to \bprp = 3 for $f > 0$ cases using the model luminosity, gravity, mass, and dual-temperature regime (for hot and cool surface regions) to obtain bolometric corrections (and hence absolute magnitudes) for \gaia\ \textit{G}, \textit{B$_{P}$}, and \textit{R$_{P}$} and 2MASS \textit{Ks} magnitudes from the YBC database of bolometric corrections \citep{Chen2019}, assuming A$_V$=0 and solar metallicity. The \bprp\ colors were generated by taking the difference between the absolute $B_{\rm P}$ and $R_{\rm P}$ magnitudes. 
 
We computed the spot fractions $f$, which are specified in terms of surface area, to fractions in luminosity $f'$ using: 
\begin{equation}
     f' = \frac{f}{ f + (1-f)(T_{\rm hot}/T_{\rm cool})^4}.
\label{eqn:fraction}
\end{equation} 
The two model temperatures, $T_{\rm hot}$ and $T_{\rm cool}$ from SPOTS were used to separately compute magnitudes and relative fluxes for hot and cool surface regions, and an average flux was computed based on our modified spot fractions. Objects redder than \bprp=3 are still included in the fits.
 
\begin{figure*}
    \centering
    \includegraphics[width=\textwidth]{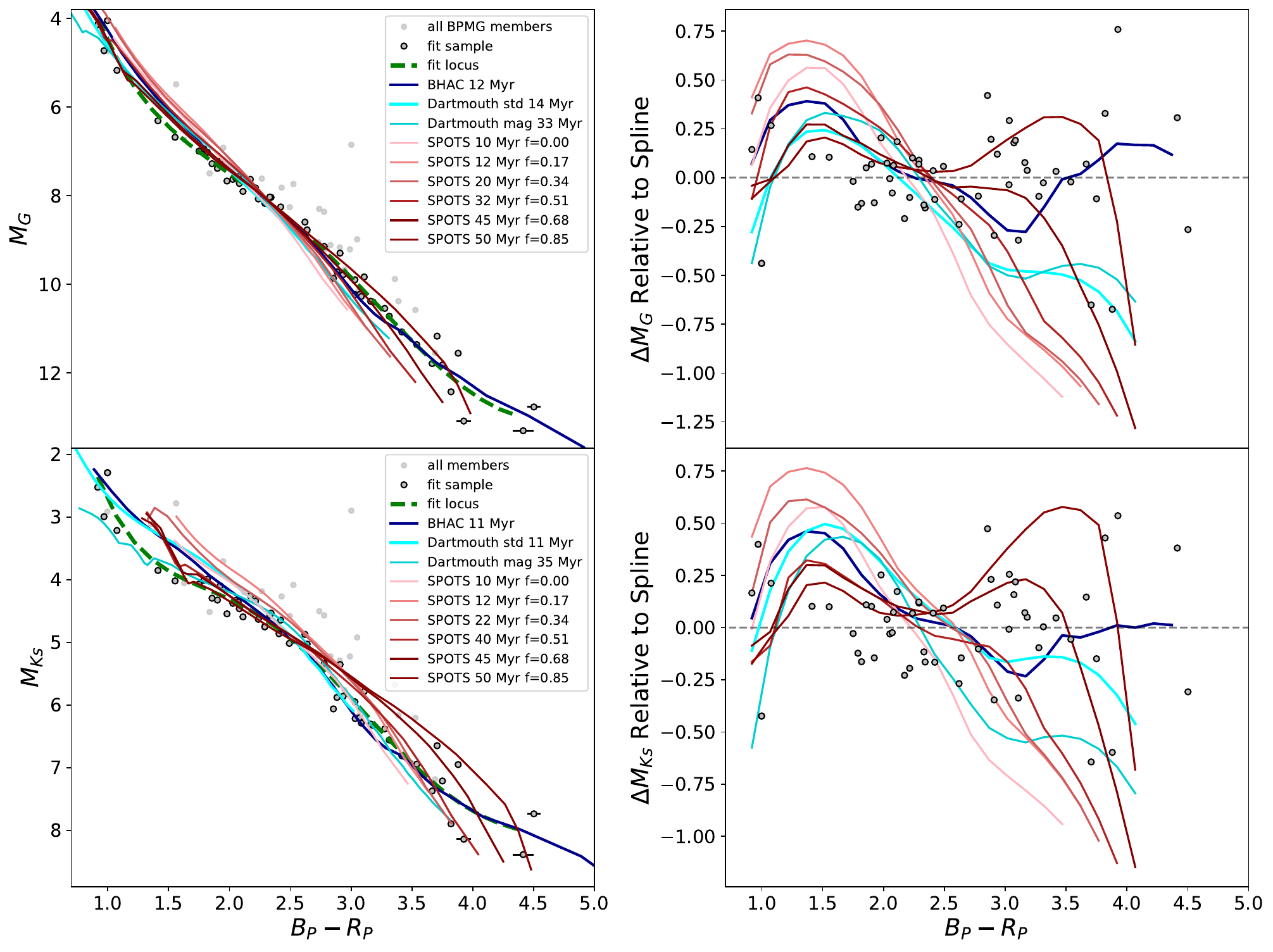}
   \caption{\gaia\ (top) and 2MASS-\gaia\ (bottom) color-magnitude diagrams of the single star BPMG sample. All BPMG members confirmed in this study are shown as grey dots, while those passing the binary exclusion and photometric criteria are outlined in black. Only those outlined are used in the isochrone fitting. \textit{Left}: Best-fit isochrones by minimum $\tau^2$. For the SPOTS isochrones, \textit{f $^\prime$} denotes the revised spot fraction (Eq. \ref{eqn:prob}) from YBC corrections. In both panels, we show a best-fit spline to the main BPMG locus in red. \textit{Right}: Each of the nine best-fit isochrone curves relative to the spline fit to better visualize the regions of the CMD where the models tend to fit or fail. The $\tau^2$ values for each fit are given in Table \ref{tab:ages}.}
\label{fig:cmd_fits}
\end{figure*}

To fit isochrones we used the $\tau^2$ metric \citep{Naylor2006}, a form of 2-D maximum-likelihood fitting. $\tau^2$ behaves essentially as a $\chi^2$ statistic, with an added dependence on the model density of stars in color-magnitude space \citep[$\rho$ in Eqn. 5 of][]{Naylor2006}.  We show in Fig. \ref{fig:cmd_fits} the results of our isochrone fitting analysis for each of the nine models (BHAC, Dartmouth standard, Dartmouth magnetic, and SPOTS models for each of six values of spot fraction), in \gaia\ \bprp\ vs.  \textit{$M_G$} and \textit{$M_{Ks}$} CMDs. To better visualize the details of model performance, the difference in absolute magnitude vs. \bprp\ for each model relative to the 5th degree polynomial fits to the main BPMG locus are plotted in Fig. \ref{fig:cmd_fits}.  (See Fig. \ref{fig:cmds} for an uncluttered look at the sample and polynomial fits). We report the best-fit ages and $\tau^2$ values from each of the fits in Table \ref{tab:ages}.

\subsection{Lithium Depletion Boundary Analysis}
\label{sec:ldb}
We compared observed and model abundance of lithium vs. effective temperature as an independent constraint on the age of the BPMG.  Measurements of the equivalent width (EW) as well as upper limits for the strength of the Li~I doublet at 6708\AA\ were gathered from the literature \citep{Bowler2019,Mentuch2008,Messina2016,Zuckerman2001} for single BPMG members. For this analysis, we chose to exclude binary systems due to confusion in deriving accurate abundances for each binary component, as well as the potential effect of a companion on rotational spin-down and Li depletion timescales compared to single stars \citep{Somers2015,Bouvier2018}.   EWs were converted to normalized abundances based on the curves of growth of \citet{Palla2007} in the \teff\ range 3100 --3600 K and \citet{Soderblom1993} in the \teff\ range 4000 -- 6500 K. We scaled these nomralized abundances to absolute values assuming A(Li) = 3.30 dex on an A(H)=12 scale, a value based on meteoritic abundances, a proxy for Li in the protosolar value \citep{Asplund2009}.  A typical abundance uncertainty was computed from the typical EW error and \teff\ error added in quadrature. We show the converted Li abundances as a function of \gaia\ \bprp, a proxy for \teff, in Fig. \ref{fig:ldb_fits}, along with predicted abundances from each of the nine models. For each of the models, we identified the best-fit models by minimum $\chi^2$ fit.

\section{Results}
\label{sec:results}

\subsection{Membership and Multiplicity}
\label{sec:membership}
 
From an initial list of 238 stars with complete kinematic information we identified 148 as members of BPMG with a Bayesian probability P $> 0.9$.  Fifteen members were already identified in the literature as binaries. 30 systems were identified as resolved binary systems in \gaia\ DR3 and of these, 14 new companions were independently confirmed in {\tt BANYAN}.  16 binary companions were resolved in AO images; 4 of which were identified in the \gaia\ search. 
Table \ref{tab:members} catalogs confirmed BPMG members in 132 systems, including 64 single stars, 42 resolved binaries, and 47 unresolved binaries. Members of binary systems and those for which the 2MASS \textit{Ks} magnitudes were recomputed based on \gaia\ photometry are flagged in Table \ref{tab:members}.

\begin{figure}
    \centering
    \includegraphics[width=\columnwidth]{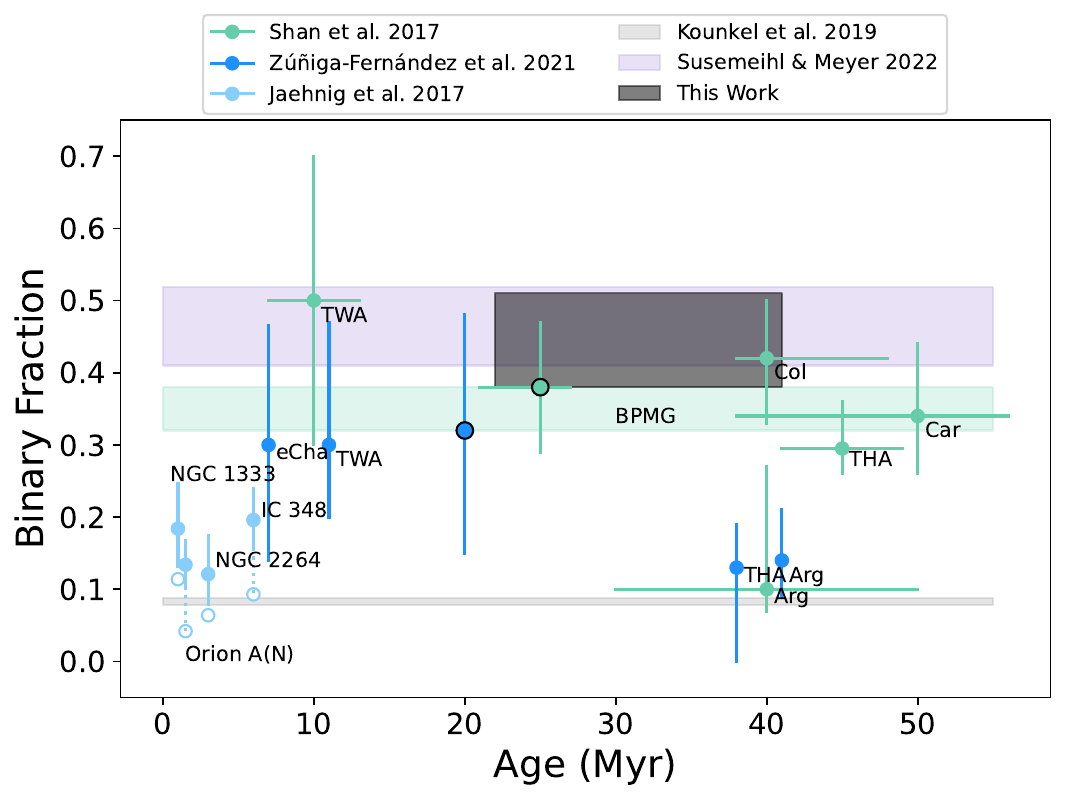}
    \caption{Binary fractions from this work compared to those from previous BPMG studies and different YMGs. Estimates for BPMG are outlined in black, and the dark grey shaded region is the result from this work. The lower bound represents the conservative binary fraction that does not include photometric binaries, and the upper bound is the fraction including photometric binaries. The \citet{Kounkel2019} (light grey shaded) region is an estimate of from RVs and the double-lined spectroscopic binary (SB2) fraction. \citet{Jaehnig2017} give spectroscopic binary (SB) fractions in star-forming regions (sky blue points). The open and closed points represent the raw observed binary fractions and modeled binary fractions correcting for completeness, respectively. \citet{Shan2017} report multiplicity fractions by association (turquoise points) as well as an overall fraction (turquoise shaded region) based on AO imaging. \citet{Zuniga2021} also report a SB fraction (dark blue points). \citet{Susemiehl2022} report a multiplicity fraction for M dwarfs, considering systems with mass ratios from 0.1 to 1, and separations from 0 to $\infty$ (purple shaded region).  \label{frac}}
\label{fig:ymg_binaries}
\end{figure}

We found that 38--51\% of BPMG systems are binaries, the proportion of binary to single systems in this sample, consistent with previous estimates for pre-MS stars in the literature  ($\sim$50-58\%, \citet{Ghez1997}). The upper bound of the multiplicity fraction fraction assumes the AO-resolved systems within 4" are bound, as is supported statistically \citep{Kraus2008}. We also include all over-luminous sources (candidate binaries identified by photometry) in this estimate, as a majority of them were additionally suggested to be binaries by \gaia\ RUWE or proper motion matching. Excluding these systems would yield a conservative multiplicity lower-limit of 38\%. We compare our results to other YMGs and the field in Fig. \ref{fig:ymg_binaries}, adapted from \citet{Shan2017} and updated with \citet{Jaehnig2017,Kounkel2019,Zuniga2021}, with age estimates based on BANYAN \citep{Gagne2018a} and SACY \citep[Search for associations containing young stars;][]{Zuniga-fernandez2020} membership catalogs.  The binary fraction in 1--50 Myr clusters and groups other than BPMG appear to have relatively lower binary fractions \citep{Jaehnig2017,Shan2017,Zuniga2021}. This is possibly because the binary analysis in this work was more thorough than those of those other, more distant groups and clusters.

\subsection{Age Estimates}
\label{sec:age_results}

\begin{figure}
    \centering
    \includegraphics[width=\columnwidth]{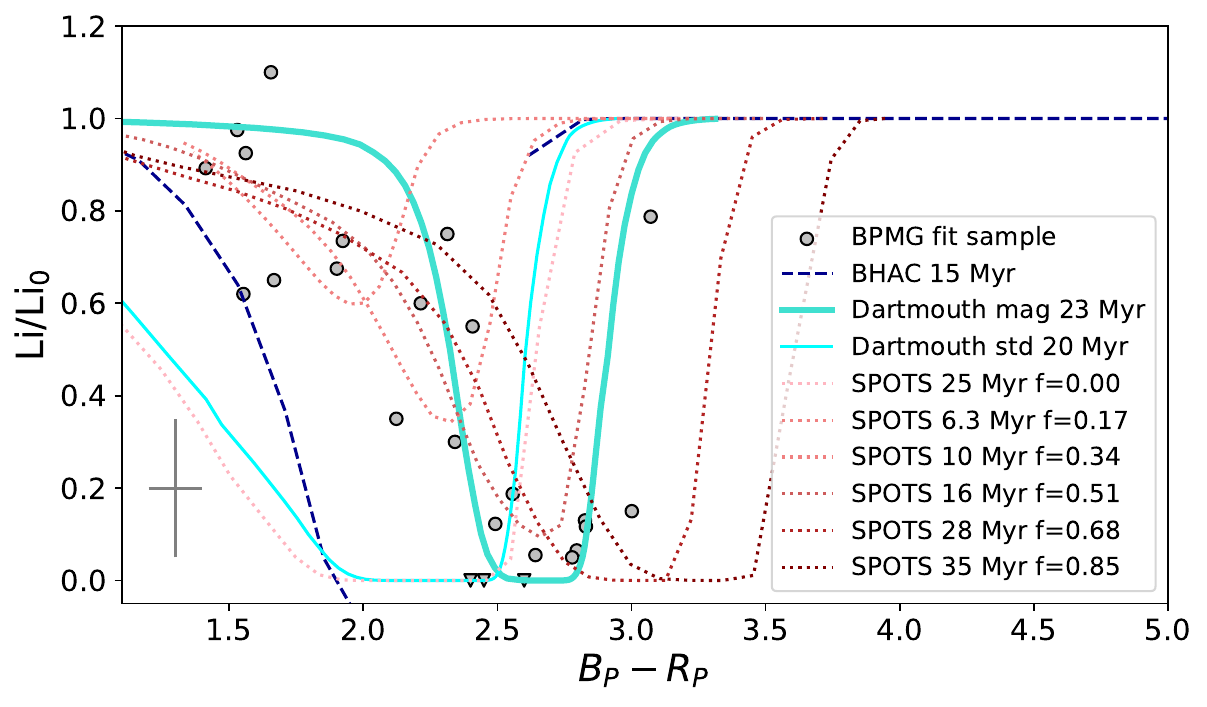}
   \caption{Li I abundances  (normalized relative to an assumed primordial value $A(Li) = 3.2$) vs. \gaia\ \bprp\ color for BPMG members with 6708\AA\ doublet equivalent width measurements from the literature. Points plotted as triangles represent upper limits. The typical measurement uncertainty is represented as the error bar in the bottom left.  For each model, the best (minumum $\chi^2$) fit model LDB is plotted.  For the incomplete LDB curves from the \citet{Baraffe2015} (BHAC) models, the fitting sample includes only those within the color ranges covered by the models. }
\label{fig:ldb_fits}
\end{figure}

\begin{figure}
    \centering
    \includegraphics[width=\columnwidth]{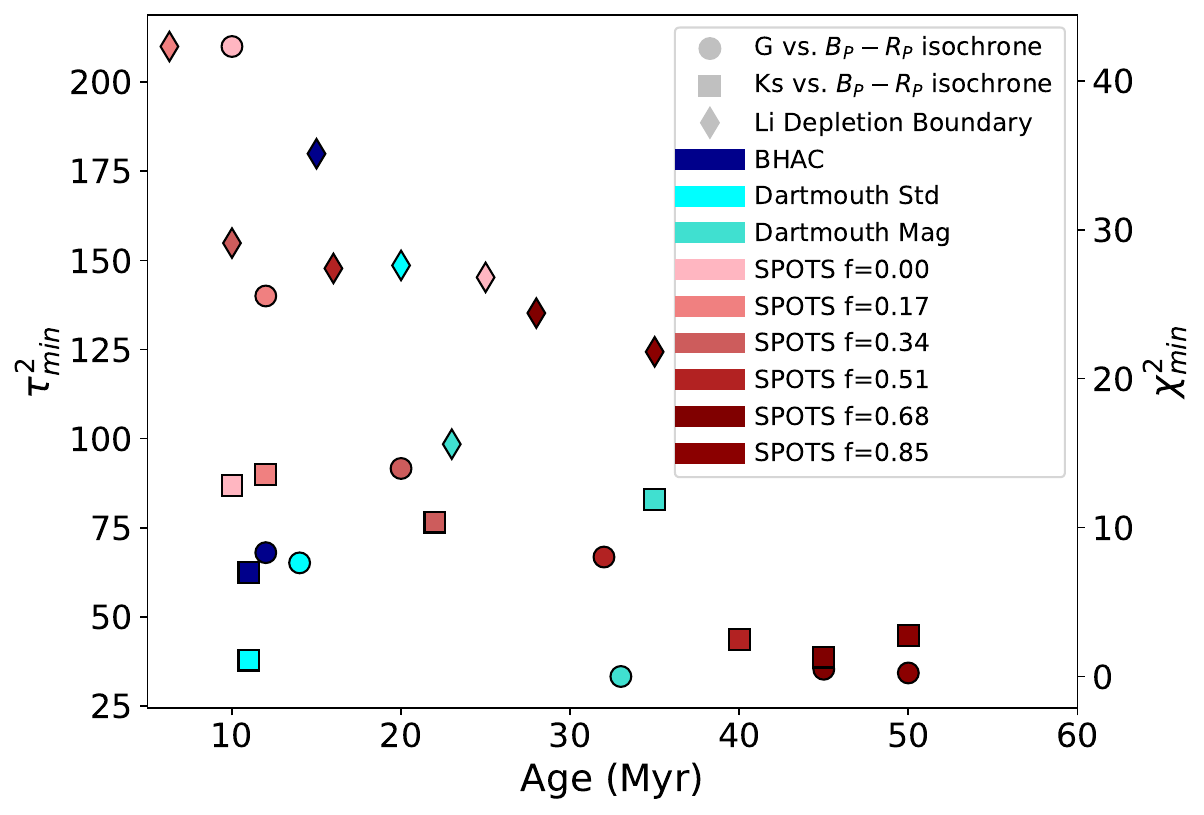}
\caption{The results in Table \ref{tab:ages} shown graphically. For CMD fits (circles, squares) the y-axis is $\tau^2$ scaled on the left, while for LDB fits (diamonds), the y-axis is $\chi^2$ scaled on the right. The two y-axis scales are not related.}
    \label{fig:age_estimates}
\end{figure}

For our age analysis, we excluded 47 confirmed members systems due to unresolved binarity; 29 of these were based on the \gaia\ RUWE criterion, 21 based on RV variability, and 18 based on photometry; with some systems flagged by multiple methods.  We show the best-fit isochrones resulting from the CMD fits in Fig. \ref{fig:cmd_fits}.  The LDB fits are shown in Fig. \ref{fig:ldb_fits}. We tabulate the best-fit values from each model and fit method in Table \ref{tab:ages}, and these are shown graphically in Fig. \ref{fig:age_estimates}.  For the Dartmouth Standard and Magnetic models, from which the lowest fit residuals for each method were derived, we assessed the concordance or internal agreement among the ages derived from the three fit methods (\gaia-CMD, \gaia-2MASS-CMD, and LDB).  We performed an ANOVA analysis to assess the variance between age estimates.  To do this, we computed an F-statistic as the ratio of inter-method to intra-method $\chi^2$. This analysis should be considered a rough representation of the method concordance, since we do not compute individual ages for every star, but rather a single age fit to the group locus. The measurements for each method are therefore independent of each other, and their variances cannot necessarily be compared in the way an ANOVA intends. The Dartmouth Magnetic models are the best-fitting in both the \gaia\ $M_G$ vs. \bprp CMD and LDB. The method concordance (F=0.34; p=0.97; $
alpha$=0.05) suggests, however, that the internal agreement between ages is not significant, and the variance is likely attributed to the differences in the models as well as the scatter in the data. We find a similar result for the Dartmouth Standard model fits (F=0.29; p=0.97; $
alpha$=0.05).

\section{Discussion}
\label{sec:discussion}


We have constructed an updated catalog of the Beta Pictoris Moving Group using the most precise available photometry and astrometry and utilizing complete kinematics for membership validation. We further refined this sample by quantitatively identifying older MS interlopers aligned in kinematic space with the BPMG, but exhibiting older ages on a color-magnitude diagram. Our catalog contains a total of 106 single and resolved companions, and 47 unresolved binaries. To date, this is the most thorough vetting of stellar multiplicity in the BPMG sample, and improves upon the previous samples of \citet{Gagne2018b} and subsequent literature by newly confirming 37 previously identified candidates and rejecting 90 others. The multiplicity fraction of 38--51$\pm$7\% (excluding/including photometric binaries) reported here is in agreement with the literature \citep[e.g.,][]{Shan2017,Zuniga2021} within errors (Fig. \ref{frac}). The slightly higher fraction reflects improved completeness of the binary search, as previous works have tended to utilize a subset of the binary identification methods used in this work (i.e., RVs, AO imaging). The binary fractions from \citet{Kounkel2019} and \citet{Jaehnig2017} in Fig. \ref{frac} are the ratio of confirmed spectroscopic binaries (SBs) to the total of their sample. \citet{Zuniga2021} also report SB fractions, but extrapolate to an expected occurrence rate. Since we have incompletely investigated SBs in this work (SB1s from RVs, and SB2s from literature), our multiplicity fraction is likely an underestimate.

This sample refinement appears to affect the age estimates quite variably among the different models. Curiously, with the refined single-star fit sample, one would expect the isochrone-based ages to bias older \citep{Sullivan2021}, but the BHAC and Dartmouth Standard models consistently produce younger ages ($\sim$11-20 Myr) compared to the literature ($\sim$15-28 Myr) employing the same models \citep[e.g.,][]{Binks2014,Malo2014b}. One key difference between the current and previous studies is the availability of the latest \gaia\ parallaxes to compute absolute magnitudes. 

\begin{table*}
\caption[Results of model fitting]{BPMG age estimates for each of the 9 evolutionary model sets and 3 age-dating methods (2 CMD and 1 LDB). The 1$\sigma$ errors are given in the Age column for the best-fitting models (Dartmouth Standard and Magnetic; see Sec. \ref{sec:age_results}). Minimum $\tau^2$ values are given for CMD best-fit isochrones (see Sec. \ref{sec:cmd}). Additionally, the minimum $\chi^2$ value from the best-fit LDB ages. The best-fit age results with lowest overall residuals from each fit method (\textit{Gaia} CMD, \textit{Gaia}-2MASS CMD, and LDB) are boldface.
   \label{tab:ages}}
\vspace{4ex}
\begin{center}
\small
\begin{tabular}{|l|c|c|c|c|}
\hline
Model & Method & Age (Myr) & $\tau^2_{min, fit}\;(\nu=1)$ & $\chi^2_{min, fit}\;(\nu=1)$  \\
\hline
BHAC          &\textit{G} vs. B$_{\rm P}$-R$_{\rm P}$ CMD  &12   & 68.0 & --  \\
BHAC          &\textit{Ks} vs. B$_{\rm P}$-R$_{\rm P}$ CMD  &11   &62.5 & -- \\
BHAC          &LDB  & 15   & -- & 35.1 \\
\hline
Dartmouth Standard         &\textit{G} vs. B$_{\rm P}$-R$_{\rm P}$ CMD  &14$^{+6}_{-4}$   &65.1  &  --  \\
\textbf{Dartmouth Standard }         &\textbf{\textit{Ks} vs. B$_{\rm P}$-R$_{\rm P}$ CMD}  &\textbf{11$^{+4}_{-3}$}   &\textbf{37.8} & --  \\
Dartmouth Standard          &LDB  & 20$\pm5$   &  -- & 27.6  \\
\hline
\textbf{Dartmouth Magnetic}         &\textbf{\textit{G} vs. B$_{\rm P}$-R$_{\rm P}$ CMD}  &\textbf{33$^{+9}_{-11}$}   &\textbf{33.2} & --  \\
Dartmouth Magnetic          &\textit{Ks} vs. B$_{\rm P}$-R$_{\rm P}$ CMD  &35$^{+10}_{-8}$   &82.9 & --  \\
\textbf{Dartmouth Magnetic }         &\textbf{LDB } &\textbf{23$\pm 8$}   &  -- & \textbf{15.6}  \\
\hline
SPOTS f=0.00         &\textit{G} vs. B$_{\rm P}$-R$_{\rm P}$ CMD  &10   &210.0  & --  \\
SPOTS f=0.00          &\textit{Ks} vs. B$_{\rm P}$-R$_{\rm P}$ CMD  &10   &86.7 & --  \\
SPOTS f=0.00          &LDB  &25   & -- & 26.8  \\
\hline
SPOTS f=0.17         &\textit{G} vs. B$_{\rm P}$-R$_{\rm P}$ CMD  &12   &140.0 & --  \\
SPOTS f=0.17          &\textit{Ks} vs. B$_{\rm P}$-R$_{\rm P}$ CMD  &12   &89.8 & -- \\
SPOTS f=0.17          &LDB  &6.3   & -- & 42.3  \\
\hline
SPOTS f=0.34         &\textit{G} vs. B$_{\rm P}$-R$_{\rm P}$ CMD  &20   &91.6 & --  \\
SPOTS f=0.34          &\textit{Ks} vs. B$_{\rm P}$-R$_{\rm P}$ CMD  &22   &76.6  & --   \\
SPOTS f=0.34          &LDB  &10   & -- &  29.1  \\
\hline
SPOTS f=0.51        &\textit{G} vs. B$_{\rm P}$-R$_{\rm P}$ CMD  &32   &66.8 & --  \\
SPOTS f=0.51          &\textit{Ks} vs. B$_{\rm P}$-R$_{\rm P}$ CMD  &40   &43.5 & --  \\
SPOTS f=0.51          &LDB  &16   & -- & 27.4  \\
\hline
SPOTS f=0.68         &\textit{G} vs. B$_{\rm P}$-R$_{\rm P}$ CMD  &45   &35.3 & --  \\
SPOTS f=0.68          &\textit{Ks} vs. B$_{\rm P}$-R$_{\rm P}$ CMD  &45   &38.7 & --  \\
SPOTS f=0.68          &LDB  &28   & -- &  24.4  \\
\hline
SPOTS f=0.85         &\textit{G} vs. B$_{\rm P}$-R$_{\rm P}$ CMD  &50   &34.2  & --  \\
SPOTS f=0.85          &\textit{Ks} vs. B$_{\rm P}$-R$_{\rm P}$ CMD  &50   &44.7 & --  \\
SPOTS f=0.85          &LDB  &35   & -- & 28.1  \\

\hline
\end{tabular}
\end{center}
\end{table*}

There is notable internal disagreement between the models used in this study, suggesting that such age determinations are largely model- (and method-) dependent. We show in Fig. \ref{fig:bpmg_ages} the results of the present work in comparison to previous literature values. The new age estimates from the non-standard models still lie within the uncertainty spread of the literature, but tend to lie at the older end of the range. The overall best fits (by minimum $\tau^2$ or $\chi^2$, producing the lowest fit residuals) from each method (the two CMDs and LDB) from the 27 estimates are in boldface in Table \ref{tab:ages}.   

The model age estimate with the lowest overall residual is given by the Dartmouth magnetic models, which yield the lowest $\tau^2$ value in a CMD. The Dartmouth magnetic models also result in the lowest fit residuals in two of the three fit methods. The Gaia \textit{G} vs. \bprp\ CMD places the BPMG at 33 Myr according to the \gaia\ CMD, though the best-fit LDB suggests 23 Myr. We assign an uncertainty on the BPMG age due to model selection of about $\pm$10 Myr (see Fig. \ref{fig:bpmg_ages}). The overall best-fit \gaia-2MASS isochrone is from the Dartmouth standard model and suggests a much younger age of 11 Myr, but the magnetic model returns a slightly lower $\tau^2$ value. These two isochrone curves (Dartmouth magnetic 33 Myr and standard 11 Myr) are nearly identical, highlighted by the similarity in their offsets from the BPMG locus in the upper right panel of Fig. \ref{fig:cmd_fits} (teal and cyan curves). The magnetic model should be more appropriate, as it should better approximate the actual physics in these highly active stars \citep{Morris2020}.

One caveat of the isochrone fitting is that not all models are appropriate for all stars. We fit the isochrone models for BPMG members spanning a large mass range for homogeneity, however, it is expected that different models are better suited for age determination in different mass ranges. While the lower mass members should be better described by the magnetic and spot fraction models, the higher mass and hotter stars are expected to have fewer spots \citep{Morris2020}. 

In the \gaia\ $M_G$ vs. B$_{\rm P}$-R$_{\rm P}$ CMD, the standard models (BHAC and Dartmouth standard) produce the shape of the BPMG locus well, but slightly overestimate the luminosity of the faintest objects.  The standard models do not reproduce the shape of the LDB observed in BPMG. This, in conjunction with the poorer isochrone fits, suggests that the magnetic and spot fraction models are better suited for age-dating BPMG and similar YMGs and young associations.  The 2MASS $K_s$ vs. B$_{\rm P}$-R$_{\rm P}$ CMD generally produced poorer isochrone fits across the 9 models. And in both CMDs, the models tend to overestimate the luminosity of the bluest stars. 

With increasing spot fraction, the SPOTS models tend to better fit the hottest stars while underestimating the luminosity or color of the coolest members. As mentioned in Sec. \ref{sec:cmd}, the SPOTS models are not well calibrated in this region of the CMD \citep{Somers2020}. It is possible that the YBC-interpolated isochrones used here are neglecting to properly account for extinction and reddening, though this should not be significant for the majority of this sample. Another possibility is that the revised luminosity spot fractions (as opposed to the surface spot fraction) overestimate the true spot fractions of the low mass members, but not the higher mass members. Hence, the spot fractions should not be considered constant with stellar mass (and T$_{\rm{eff}}$), as was assumed when the isochrones were generated. Among the SPOTS models, the overall best-performing is the f $^\prime$ = 0.837 model, which suggests and age range of 28-45 Myr. However, very high spot fractions are likely unrealistic since they have not been observed among young stars \citep{Cao2023}, and would imply lower photometric variability than observed. 

Fitting to the CMD vs. the LDB returned marginally discrepant ages of $33^{+11}_{-9}$ and $23\pm8$ Myr, respectively.  One explanation for this divergence, compared to, e.g., the consistent CMD and LDB ages in \citet{Malo2014b}, is the use of revised solar abundances from \citet{Grevesse2007} in the stellar models over previous estimates \citep{Grevesse1998}. It is possible that the \citet{Grevesse2007} abundances used in the current models bias CMD ages older compared to those from \citep{Grevesse1998}. However, the precise effects of solar abundance estimates on model ages for young stars, and which abundances are to be adopted, are yet to be determined \citep[e.g.,][]{Asplund2021,Serenelli2009}.  Another explanation could be a non-solar metallicity for the BPMG, but the deployment of solar-metallicity models.  \citet{VianaAlmeida2009} estimated the metallicity of the BPMG as [Fe/H] = $-0.01\pm 0.08$, but only from a single star.  Preliminary tests with the Dartmouth magnetic models (G. Feiden, pers. comm.) show that a super-solar metallicity like that found in nearby young clusters \citep{Brandt2016,Dutra-Ferreira2016} shifts the lithium depletion pattern to higher masses, but lower \teff, at a given age. This could decrease or increase the age estimate depending on the relative sampling of the warmer vs. cooler sides of the boundary (Fig. \ref{fig:ldb_fits}).  These models also show that a higher metallicity produces a systematic shift to older ages in a CMD of these fully convective stars.  Clearly, a robust estimate for the metallicity of the BPMG is needed and, if necessary, fitting with additional models appropriate to that value.

It is important to also consider the connections between multiplicity, rotation, and age estimates \citep{Messina2016,Messina2019}.  Tidal locking in close binaries can lead to spin-up as angular momentum is transferred from the orbit to rotation, while widely separated companions can enable spin-up by truncating and accelerating the dissipation of the circumstellar disk, halting angular momentum transfer from the star \citep{Rosotti2017}.  Both cases lead to more rapid rotation, but past studies of the connections between binary separation, disk dissipation, and rotation have been ambiguous or contradictory \citep[e.g.,][]{Messina2016,Allen2017,Kuruwita2018}. Rotation affects the magnetic and surface activity, complicating isochrone fitting, but also affects interior mixing and lithium depletion \citep{Somers2015,Messina2016,Bouvier2018}. \citet{Messina2016} employed the same Dartmouth models for age-dating BPMG and reported that the results significantly improve when Li EW is de-trended with rotation.

If an older age for BPMG is borne out, there are several implications for our understanding of disk and planet evolution. If the BPMG is indeed older than previously accepted, this is in agreement with the expected typical disk lifetimes of a few to tens of Myr \citep{Richert2018}, since only debris disks remain in BPMG. Additionally, any wide-orbit, directly detected substellar and planetary companions of BPMG members, e.g. 51 Eridani, and other well-studied systems mentioned in Sec. \ref{sec:intro}, would be more massive than previously determined based on their luminosity \citep{Sullivan2021}.

The relationship between binary separation, formation, and pre-MS evolution is not well known; this sample of young pre-MS binaries at varying separations paves the way for a detailed study of how the presence of a companion may affect pre-MS spin-down, especially since much of BPMG has been observed by continuous photometric surveys such as TESS.

\section{Conclusions}

In this study we presented an updated membership census and binary catalog for the Beta Pictoris Moving Group, the nearest and youngest co-moving stellar group in the solar neighborhood. We utilized the latest and most precise space-based astrometry, photometry, and radial velocity data from \gaia\ DR3 to compute precise membership probabilities and identify binaries contaminating the samples of previous photometric age-dating studies. The \gaia\ data was supplemented with ground-based RVs from surveys and literature, and photometry from the 2MASS survey. 

With the \gaia\ and RV survey data, we were able to confirm 37 new members of BPMG and reject 90 candidates (Table \ref{tab:rejects}) previously identified in the literature which lacked full kinematic data. We found multiplicity fraction in the range 38--51$\pm$7\% using multiple methods to identify resolved and unresolved binaries among the confirmed BPMG members. The single-star and resolved companion sample was used in age-dating analyses to minimize the expected age-biasing effects of unidentified binaries in a stellar sample.  

We used 9 suites of evolutionary models: the BHAC standard models, Dartmouth standard and magnetic models, and the SPOTS models at 6 discrete spot fractions. We found that the difference in isochrone age between our refined sample and earlier works is smaller than the variation among isochrone ages for the different models used here. From isochrone fitting in two CMDs and from LDB fitting, we found the BPMG age ranges from 10-50 Myr. This result  highlights the sensitivity of age estimates to different magnetic and spot fraction models. However, the same standard models employed in previous works still give much younger ages of 10-12 Myr, which indicates that unresolved binaries have not had a major effect on previous determinations of BPMG's age.  This study shows that isochrone fitting for young stellar populations is largely sensitive to the models themselves, and that, overall, the surface spot and magnetic models out-perform the previous standards. 

\section*{Acknowledgements}

We thank Gary Huss for considerable and constructive feedback on earlier versions of this manuscript.  R.A.L., E.G. and J. vS. are supported by NSF Astronomy \& Astrophysics Research Grant 1817215.  We thank the anonymous reviewer for insightful comments which have greatly improved the quality of this manuscript.  This research has made use of the Keck Observatory Archive (KOA), which is operated by the W. M. Keck Observatory and the NASA Exoplanet Science Institute (NExScI), under contract with the National Aeronautics and Space Administration. This work has made use of data from the European Space Agency (ESA) mission \gaia\ (\url{https://www.cosmos.esa.int/gaia}), processed by the {\it Gaia} Data Processing and Analysis Consortium (DPAC, \url{https://www.cosmos.esa.int/web/gaia/dpac/consortium}).  Funding for the DPAC has been provided by national institutions, in particular the institutions participating in the \gaia\ Multilateral Agreement. This publication makes use of data products from the Two Micron All Sky Survey, which is a joint project of the University of Massachusetts and the Infrared Processing and Analysis Center/California Institute of Technology, funded by the National Aeronautics and Space Administration and the National Science Foundation. Based on observations obtained at the international Gemini Observatory, a program of NSF’s NOIRLab, which is managed by the Association of Universities for Research in Astronomy (AURA) under a cooperative agreement with the National Science Foundation on behalf of the Gemini Observatory partnership: the National Science Foundation (United States), National Research Council (Canada), Agencia Nacional de Investigaci\'{o}n y Desarrollo (Chile), Ministerio de Ciencia, Tecnología e Innovaci\'{o}n (Argentina), Minist\'{e}rio da Ci\'{e}ncia, Tecnologia, Inovaç\~{o}es e Comunicaç\~{o}es (Brazil), and Korea Astronomy and Space Science Institute (Republic of Korea).  Based on data obtained from the ESO Science Archive Facility. This work has made use of open source Python packages {\tt NumPy} \citep{Numpy} and {\tt SciPy} \citep{Scipy2019}.

\section*{Data Availability}

All data used for this work are either available from the public archives (CDS) or on request from the authors.  The modified SPOTS+YBC isochrones are available on Zenodo at: [LINK TO BE PROVIDED]

\bibliographystyle{mnras}
\bibliography{references_master,references_extra}




\section*{Appendix A:}
We list the confirmed BPMG members and their binary designation in Table \ref{tab:members}, and provide a list of rejected candidates in Table \ref{tab:rejects}. 

\begin{landscape}
\begin{table}
\caption{Confirmed members of the Beta Pictoris Moving Group$^{1}$}
\label{tab:members}
\vspace{1ex}
\noindent
\begin{center}
\begin{tabular}{|r|r|l|c|c|c|c|c|c|c|}
\hline
\multicolumn{1}{c}{\gaia\ DR3 ID} & \multicolumn{1}{c}{2MASS ID} & \multicolumn{1}{c}{Common Name} &  \multicolumn{1}{c}{RA (hh:mm:ss.ss)} &  \multicolumn{1}{c}{Dec (dd:mm:ss.s)} & \multicolumn{1}{c}{$\pi$ (mas)} &    \multicolumn{1}{c}{\gaia\ $G$} &  \multicolumn{1}{c}{\bprp} & \multicolumn{1}{c}{Binary$^{2}$} &  \multicolumn{1}{c}{Separation ($\prime\prime$)$^{3}$} \\
\hline
\hline
2315841869173294208 & 00275035-3233238 &        GJ  2006  B & 00:27:50.35 & -32:33:24.13 &      28.59 & 11.92 &   2.80 &       R &    17.890 \\
2315849737553379840 & 00281434-3227556 &            GR* 9 & 00:28:14.35 & -32:27:55.62 &      28.56 & 13.44 &   3.31 &       N &         \\
2357025657739386624 & 00482667-1847204 &                & 00:48:26.67 & -18:47:20.42 &      19.42 & 13.94 &   3.16 &       N &         \\
2581708281894978176 & 00501752+0837341 &                & 00:50:17.51 & +08:37:34.20 &       6.31 & 12.85 &   3.00 &       U &         \\
2788357364871430400 & 01025097+1856543 &                & 01:02:50.99 & +18:56:54.16 &      26.19 & 12.58 &   2.95 &       U &         \\
2477870708709917568 & 01351393-0712517 &     Barta  161  12 & 01:35:13.92 & -07:12:51.46 &      26.82 & 11.97 &   2.83 &       U &         \\
73034991155555456 & 02175601+1225266 &                & 02:17:56.02 & +12:25:26.42 &      15.90 & 12.81 &   2.64 &       U &         \\
87555176071871744 & 02241739+2031513 &                & 02:24:17.40 & +20:31:51.42 &      14.13 & 16.04 &   3.67 &       N &         \\
132363027978672000 & 02272804+3058405 &         AG  Tri  B & 02:27:28.06 & +30:58:40.36 &      24.43 & 11.43 &   2.41 &       R &    22.050 \\
132362959259196032 & 02272924+3058246 &         AG  Tri  A & 02:27:29.25 & +30:58:24.60 &      24.42 &  9.74 &   1.55 &       U &         \\
22338644598132096 & 02442137+1057411 &          MCC  401 & 02:44:21.37 & +10:57:41.07 &      20.78 & 10.33 &   1.95 &       R &     0.223 \\
5177677603263978880 & 02450826-0708120 &                & 02:45:08.27 & -07:08:12.09 &      14.66 & 14.00 &   3.11 &       N &         \\
5183875103632956032 & 02495639-0557352 &                & 02:49:56.38 & -05:57:35.43 &      15.09 & 15.66 &   3.88 &       U &         \\
68012529415816832 & 03350208+2342356 &   EPIC  211046195 & 03:35:02.09 & +23:42:35.41 &      19.72 & 16.30 &   4.50 &       N &         \\
 244734765608363136 & 03393700+4531160 &                & 03:39:37.01 & +45:31:15.97 &      24.82 & 12.89 &   2.85 &       U &         \\
  66245408072670336 & 03573393+2445106 &       StKM  1-433 & 03:57:33.93 & +24:45:10.64 &      14.55 & 11.86 &   1.98 &       N &         \\
3205094369407459456 & 04373746-0229282 &          GJ  3305 & 04:37:37.46 & -02:29:28.95 &      36.01 &  9.80 &   2.12 &       R &     0.235 \\
175329120598595200 & 04435686+3723033 &  PM  J04439+3723W & 04:43:56.87 & +37:23:03.36 &      14.01 & 12.31 &   2.34 &       R &     7.650 \\
3231945508509506176 & 04593483+0147007 &        V1005  Ori & 04:59:34.83 & +01:47:00.67 &      40.99 &  9.32 &   1.90 &       N &         \\
4764027962957023104 & 05004714-5715255 &       CD-57  1054 & 05:00:47.13 & -57:15:25.45 &      37.21 &  9.38 &   1.92 &       N &         \\
3228854506445679616 & 05015665+0108429 &                & 05:01:56.65 & +01:08:42.90 &      39.54 & 11.51 &   2.86 &       U &         \\
3238965099979863296 & 05061292+0439272 &                & 05:06:12.93 & +04:39:27.18 &      36.19 & 11.99 &   2.93 &       N &         \\
2962658549474035584 & 05064991-2135091 &     BD-21  1074  A & 05:06:49.50 & -21:35:04.30 &      50.43 &  9.56 &   2.24 &       R &     7.790 \\
2901786974419551488 & 05294468-3239141 &   SCR  J0529-3239 & 05:29:44.69 & -32:39:14.30 &      33.60 & 12.27 &   3.03 &       N &         \\
3216729573251961856 & 05320450-0305291 &     V*  V1311  Ori & 05:32:04.50 & -03:05:29.40 &      27.22 & 10.44 &   2.27 &       R &     0.159 \\
3216922640621225088 & 05335981-0221325 &                & 05:33:59.82 & -02:21:32.38 &      29.12 & 11.27 &   2.49 &       U &         \\
2899492637251200512 & 06131330-2742054 &                & 06:13:13.32 & -27:42:05.57 &      29.62 & 10.94 &   2.74 &       R &     0.110 \\
5266270443442455040 & 06182824-7202416 &           AO  Men & 06:18:28.21 & -72:02:41.45 &      25.57 &  9.27 &   1.41 &       N &         \\
5412403269717562240 & 09462782-4457408 &                & 09:46:27.82 & -44:57:40.85 &      21.44 & 14.52 &   3.71 &       N &         \\
5355751581627180288 & 10172689-5354265 &         TWA  22  A & 10:17:26.89 & -53:54:26.39 &      50.52 & 12.06 &   3.53 &       R &     0.106 \\
5849837854861497856 & 14423039-6458305 &        *  alf  Cir & 14:42:30.42 & -64:58:30.49 &      60.99 &  3.17 &   0.54 &       U &         \\
5882581895192805632 & 15385679-5742190 &      V343  Nor  BC & 15:38:56.78 & -57:42:18.95 &      25.44 & 13.21 &   3.03 &       R &    10.450 \\
5882581895219921024 & 15385757-5742273 &       V343  Nor  A & 15:38:57.56 & -57:42:27.35 &      25.83 &  7.68 &   1.00 &       U &         \\
5935776714456619008 & 16572029-5343316 &                & 16:57:20.27 & -53:43:31.58 &      19.69 & 11.35 &   2.43 &       R &     0.100 \\
5963633872326630272 & 17024014-4521587 &   UCAC2  12510535 & 17:02:40.16 & -45:21:58.72 &      31.30 & 10.70 &   2.29 &       N &         \\
4107812485571331328 & 17150362-2749397 &      CD-27  11535 & 17:15:03.61 & -27:49:39.74 &      12.00 & 10.09 &   1.56 &       R &     0.141 \\
5811866422581688320 & 17172550-6657039 &      HD  155555  B & 17:17:25.51 & -66:57:03.73 &      32.95 &  6.46 &   1.00 &       U &         \\
5811866422581688320 & 17172550-6657039 &      HD  155555  A & 17:17:25.51 & -66:57:03.73 &      32.95 &  6.46 &   1.00 &       U &         \\
5811866358170877184 & 17173128-6657055 &      HD  155555  C & 17:17:31.29 & -66:57:05.47 &      32.88 & 11.47 &   2.73 &       R &    34.010 \\
5924485966955008896 & 17295506-5415487 &       CD-54  7336 & 17:29:55.08 & -54:15:48.65 &      14.79 &  9.32 &   1.08 &       N &         \\
4067828843907821824 & 17520173-2357571 & UCAC4  331-124196 & 17:52:01.74 & -23:57:57.22 &      15.67 & 11.85 &   2.21 &       N &         \\
4050178830427649024 & 18041617-3018280 &                & 18:04:16.18 & -30:18:27.96 &      18.15 & 11.74 &   2.33 &       N &         \\
\hline
\end{tabular}
\begin{quote}
$^{1}$The extended machine-readable catalog is available by request.  $^{2}$N = not binary, U = unresolved binary, and R = resolved binary.   $^{3}$For wide binaries or those resolved by \gaia\ or AO.
\end{quote}
\end{center}
\end{table}
\end{landscape}

\begin{landscape}
\begin{table}
Table~\ref{tab:members}, continued.\hfill\\

\noindent
\begin{center}
\begin{tabular}{|r|r|l|c|c|c|c|c|c|c|}
\hline
\multicolumn{1}{c}{\gaia\ DR3 ID} & \multicolumn{1}{c}{2MASS ID} & \multicolumn{1}{c}{Common Name} &  \multicolumn{1}{c}{RA (hh:mm:ss.ss)} &  \multicolumn{1}{c}{Dec (dd:mm:ss.s)} & \multicolumn{1}{c}{$\pi$ (mas)} &    \multicolumn{1}{c}{\gaia\ $G$} &  \multicolumn{1}{c}{\bprp} & \multicolumn{1}{c}{Binary$^{2}$} &  \multicolumn{1}{c}{Separation ($\prime\prime$)$^{3}$} \\
\hline
\hline
6648834361774839040 & 18055491-5704307 &  UCAC3 66-407600 & 18:05:54.92 & -57:04:30.74 &      17.71 & 12.36 &   2.62 &       N &         \\
6414282147589248000 & 18090694-7613239 &                & 18:09:06.93 & -76:13:23.89 &      36.66 & 13.26 &   3.42 &       N &         \\
6653162456161626368 & 18092970-5430532 &                & 18:09:29.71 & -54:30:53.27 &      25.67 & 13.34 &   3.39 &       U &         \\
4045698732855626624 & 18142207-3246100 &                & 18:14:22.07 & -32:46:10.13 &      13.92 & 11.91 &   2.17 &       R &     1.200 \\
6706431763001502848 & 18151564-4927472 &                & 18:15:15.64 & -49:27:47.30 &      16.13 & 11.70 &   2.53 &       U &         \\
4051081838710783232 & 18195221-2916327 &        HD 168210 & 18:19:52.21 & -29:16:32.82 &      12.43 &  8.67 &   0.92 &       U &         \\
6705107126367751168 & 18265401-4807022 &        HD 169405 & 18:26:54.01 & -48:07:02.06 &      12.20 &  5.22 &   1.03 &       U &         \\
6723183789033085824 & 18283208-4129081 &  TYC 7909-2501-1 & 18:28:32.09 & -41:29:08.36 &      14.96 & 10.86 &   1.63 &       U &         \\
6649786646225001984 & 18420483-5554126 &                & 18:42:04.84 & -55:54:12.74 &      19.36 & 13.84 &   3.08 &       N &         \\
6649788119394186112 & 18420694-5554254 &                & 18:42:06.96 & -55:54:25.58 &      19.44 & 12.33 &   2.64 &       N &         \\
6728469996119674112 & 18430597-4058047 &                & 18:43:05.97 & -40:58:04.78 &      17.10 & 12.10 &   2.50 &       U &         \\
6650036304082834560 & 18443965-5506502 &                & 18:44:39.66 & -55:06:50.09 &      11.02 & 10.24 &   1.36 &       U &         \\
6631685008336771072 & 18465255-6210366 &     Smethells 20 & 18:46:52.55 & -62:10:36.61 &      19.72 & 11.13 &   2.05 &       N &         \\
4071532308311834496 & 18471351-2808558 &                & 18:47:13.51 & -28:08:55.86 &      16.69 & 15.25 &   3.54 &       N &         \\
6736232346363422336 & 18504448-3147472 &      CD-31 16041 & 18:50:44.48 & -31:47:47.38 &      20.22 & 10.49 &   1.82 &       N &         \\
6655168686921108864 & 18530587-5010499 &        HIP 92680 & 18:53:05.87 & -50:10:49.90 &      21.16 &  8.10 &   0.97 &       N &         \\
4088823159447848064 & 19082195-1603249 &                & 19:08:21.96 & -16:03:24.82 &      14.35 & 15.77 &   3.69 &       U &         \\
6764026419748892160 & 19114467-2604085 &      CD-26 13904 & 19:11:44.67 & -26:04:08.53 &      14.77 &  9.88 &   1.40 &       R &     0.210 \\
6663346029775435264 & 19233820-4606316 &                & 19:23:38.21 & -46:06:31.64 &      14.03 & 11.21 &   1.79 &       N &         \\
6742986538895222144 & 19243494-3442392 &                & 19:24:34.95 & -34:42:39.37 &      19.40 & 12.77 &   2.99 &       U &         \\
6643851448094862592 & 19260075-5331269 &                & 19:26:00.75 & -53:31:26.97 &      20.94 & 12.57 &   2.91 &       U &         \\
6764421281858414208 & 19300396-2939322 &                & 19:30:03.96 & -29:39:32.45 &      16.60 & 13.20 &   2.91 &       N &         \\
6754492966739292928 & 19481651-2720319 &                & 19:48:16.52 & -27:20:31.94 &      15.47 & 12.21 &   2.29 &       N &         \\
6754492932379552896 & 19481703-2720334 &                & 19:48:17.05 & -27:20:33.50 &      15.34 & 17.16 &   3.93 &       N &         \\
6747467431032539008 & 19560294-3207186 &                & 19:56:02.94 & -32:07:18.66 &      19.54 & 11.91 &   2.78 &       R &    26.310 \\
6747467224874108288 & 19560438-3207376 &  TYC 7443-1102-1 & 19:56:04.37 & -32:07:37.67 &      19.49 & 10.83 &   1.86 &       N &         \\
6747106443324127488 & 20013718-3313139 &                & 20:01:37.17 & -33:13:14.01 &      16.68 & 11.46 &   2.07 &       N &         \\
6700649538727351040 & 20055640-3216591 &                & 20:05:56.41 & -32:16:59.19 &      20.18 & 11.20 &   2.08 &       N &         \\
6850555648387276544 & 20083784-2545256 &                & 20:08:37.85 & -25:45:25.71 &      17.85 & 14.15 &   3.18 &       N &         \\
6800238044930953600 & 20333759-2556521 &                & 20:33:37.60 & -25:56:52.12 &      22.90 & 13.08 &   3.36 &       U &         \\
6794047652729201024 & 20450949-3120266 &           AU Mic & 20:45:09.53 & -31:20:27.24 &     102.94 &  7.84 &   2.11 &       N &         \\
6833291426043854976 & 21100461-1920302 &                & 21:10:04.61 & -19:20:30.44 &      29.77 & 11.62 &   3.05 &       R &     0.141 \\
6833292181958100224 & 21100535-1919573 &                & 21:10:05.36 & -19:19:57.61 &      30.90 & 10.81 &   2.42 &       N &         \\
6801191424589717888 & 21103096-2710513 &          BRG 32B & 21:10:30.97 & -27:10:51.53 &      24.84 & 14.85 &   3.75 &       N &         \\
6801191355870240768 & 21103147-2710578 &                & 21:10:31.48 & -27:10:58.01 &      24.76 & 13.58 &   3.28 &       N &         \\
6835588645136005504 & 21200779-1645475 &                & 21:20:07.81 & -16:45:47.80 &      20.72 & 13.14 &   2.88 &       N &         \\
6400160947954197888 & 21212873-6655063 &                & 21:21:28.72 & -66:55:06.27 &      31.70 &  9.99 &   1.84 &       U &         \\
2727844441062478464 & 22085034+1144131 &                & 22:08:50.34 & +11:44:13.22 &      26.75 & 13.02 &   3.06 &       U &         \\
6608255235884536320 & 22334687-2950101 &                & 22:33:46.89 & -29:50:10.22 &      19.19 & 16.89 &   4.41 &       N &         \\
6382640367603744128 & 22424896-7142211 &                & 22:42:48.93 & -71:42:21.20 &      27.23 &  9.82 &   1.75 &       N &         \\
6603693881832177792 & 22445794-3315015 &     HIP 112312 A & 22:44:57.96 & -33:15:01.74 &      47.92 & 10.74 &   2.78 &       R &    35.880 \\
6603693808817829760 & 22450004-3315258 &     HIP 112312 B & 22:45:00.06 & -33:15:26.09 &      48.00 & 11.82 &   3.07 &       N &         \\
2433191886212246784 & 23323085-1215513 &       BD-13 6424 & 23:32:30.86 & -12:15:51.46 &      36.43 &  9.83 &   2.03 &       N &         \\
2324205785406060928 & 23355015-3401477 &                & 23:35:50.17 & -34:01:47.78 &      26.76 & 15.29 &   3.82 &       N &         \\
\hline
\end{tabular}
\begin{quote}
$^{1}$The extended machine-readable catalog is available by request.  $^{2}$N = not binary, U = unresolved binary, and R = resolved binary.   $^{3}$For wide binaries or those resolved by \gaia\ or AO. 
\end{quote}
\end{center}
\end{table}
\end{landscape}

\begin{table*}
\caption{Rejected Beta Pictoris Moving Group candidates. The probability of membership to BPMG based on proper motions, parallax, and RV is listed in column log $P_{BPMG}$.}
\label{tab:rejects}
\vspace{1ex}
\noindent
\begin{center}
\begin{tabular}{|r|r|l|c|c|c|}
\hline
\multicolumn{1}{c}{\gaia\ DR3 ID} & \multicolumn{1}{c}{2MASS ID} & \multicolumn{1}{c}{Common Name} &  \multicolumn{1}{c}{RA (hh:mm:ss.ss)} &  \multicolumn{1}{c}{Dec (dd:mm:ss.s)} & \multicolumn{1}{c}{log P$_{BPMG}$}  \\
\hline
\hline
6882840883190250752 & 20554767-1706509 &          HIP 103311 AB & 20 55 47.74 &  --17 06 51.97 &        -999 \\
1227780314970612864 & 14255593+1412101 &               J1425+1412 & 14 25 55.87 &   +14 12 09.64 &        -999 \\
4707563810327288192 & 00172353-6645124 &                  RBS 38 &  00 17 23.81 & --66 45 12.59 &        -999 \\
4630867449747185280 & 00181936-8207151 & 2MASS J00181936-8207151 &  00 18 25.97 &  --82 07 14.44 &        -999 \\
5505087564445324160 & 07063751-4943243 &               J0706-4943 &   07 06 37.46 & --49 43 23.66 &        -999 \\
2749470902772705408 & 00323480+0729271 &               J0032+0729 &  00 32 34.91 &   +07 29 25.78 & -210.166853 \\
6656461918753123968 & 19063412-5327481 &               J1906-5327 &  19 06 34.15 & --53 27 49.09 & -168.301030 \\
  81166811651514880 & 02355424+1514561 &         TYC 1216-1411-1 &  02 35 54.31 &  +15 14 54.97 & -154.970616 \\
2683714304930755712 & 22064638+0325037 &                Wolf 990 &  22 06 46.85 &    +03 24 58.90 & -110.518557 \\
2306957128025970944 & 00042123-3733083 & 2MASS J00042123-3733083 &    00 04 21.60 & --37 33 10.77 &  -96.376751 \\
3292922293081928192 & 04484752+0940281 &               HD 287167 &  04 48 47.55 &   +09 40 27.49 &  -90.142668 \\
4805600119646735616 & 05332802-4257205 &                 RBS 661 &  05 33 28.01 & --42 57 19.94 &  -82.000435 \\
 109628013733250304 & 02515408+2227299 & 2MASS J02515408+2227299 &  02 51 54.22 &  +22 27 28.22 &  -51.667562 \\
 126534203306473344 & 02383622+2614198 &               BD+25 430 &  02 38 36.29 &  +26 14 18.97 &  -43.732828 \\
6652273015676968832 & 18103569-5513435 &               J1810-5513 & 18 10 35.72 &  --55 13 45.00 &  -40.366532 \\
1704563363189801600 & 16171135+7733477 &               J1617+7733 & 16 17 11.21 &  +77 33 48.33 &  -38.022734 \\
1066495290754663040 & 10015995+6651278 &               J1001+6651 &  10 01 59.75 &  +66 51 26.43 &  -34.082494 \\
3899393459350372224 & 11515681+0731262 &               J1151+0731 & 11 51 56.69 &   +07 31 25.26 &  -33.636388 \\
5055946513425551360 & 03370343-3042318 &               J0337-3042 &   03 37 03.47 & --30 42 32.01 &  -25.756962 \\
5209118305765620096 & 08475676-7854532 &              V* EQ Cha &   08 47 56.60 & --78 54 52.78 &  -25.747147 \\
4988735051246293504 & 01132817-3821024 &               J0113-3821 &  01 13 28.31 &  --38 21 03.73 &  -25.256490 \\
6718463138927725696 & 19033299-3847058 & 2MASS J19033299-3847058 &  19 03 33.04 &  --38 47 07.85 &  -20.779892 \\
4860805567684944128 & 03255277-3601161 &               J0325-3601 &  03 25 52.82 &  --36 01 16.23 &  -20.508638 \\
5132219016567875200 & 02335984-1811525 &                  BRG 13 &   02 33 59.9 & --18 11 52.99 &  -20.089376 \\
1725580076982641280 & 14124993+8401311 &               J1412+8401 & 14 12 49.47 &   +84 01 31.83 &  -19.049149 \\
4980384088633481216 & 00325584-4405058 &               J0032-4405 &  00 32 56.03 &   --44 05 07.26 &  -18.756962 \\
4093006560668685568 & 18235116-1930168 &               HD 313300 & 18 23 51.21 & --19 30 17.49 &  -15.991400 \\
6200370648577945216 & 15063505-3639297 &               J1506-3639 &  15 06 35.02 & --36 39 30.16 &  -15.818156 \\
5050451154310013056 & 02505959-3409050 &               J0250-3409 &  02 50 59.73 &    --34 09 05.50 &  -15.782516 \\
1930104018733997312 & 23010610+4002360 &               J2301+4002 &    23 01 06.20 &   +40 02 35.79 &  -15.059982 \\
2517397846786452224 & 02261625+0617331 &                HD 15115 &  02 26 16.34 &   +06 17 32.41 &  -14.882729 \\
6701528082880793088 & 18091388-5302176 &               J1809-5302 &  18 09 13.87 &  --53 02 18.38 &  -14.785156 \\
4721078629298085760 & 02564708-6343027 &               J0256-6343 &  02 56 47.23 &  --63 43 02.57 &  -14.249492 \\
5200035927402217088 & 11493184-7851011 &              V* DZ Cha & 11 49 31.63 &  --78 51 01.12 &  -13.725842 \\
5084713036143375744 & 03363144-2619578 &               J0336-2619 &  03 36 31.55 & --26 19 58.34 &  -13.353596 \\
3920498104009539456 & 12115308+1249135 &               J1211+1249 & 12 11 53.01 &  +12 49 13.02 &  -12.404504 \\
4946632929953255040 & 02304623-4343493 &        UCAC4 232-002301 &  02 30 46.36 & --43 43 49.74 &  -12.164944 \\
 735219962087569792 & 10355725+2853316 &               J1035+2853 & 10 35 57.13 &  +28 53 30.38 &  -12.152427 \\
4920330653311504256 & 00413538-5621127 &               J0041-5621 &   00 41 35.60 & --56 21 13.79 &  -12.063486 \\
3193496682100879872 & 03550477-1032415 &               J0355-1032 &   03 55 04.85 & --10 32 42.13 &  -11.596879 \\
5050649616159457280 & 02485260-3404246 &           J0248-3404 AB &  02 48 52.73 &   --34 04 25.10 &  -11.301030 \\
1747664867537511424 & 20395460+0620118 &              BD+05 4576 &  20 39 54.70 &   +06 20 10.27 &  -11.253366 \\
1851787179879007232 & 21183375+3014346 &          TYC 2703-706-1 & 21 18 33.83 &  +30 14 34.27 &  -11.195861 \\
2851052759133803776 & 23512227+2344207 &               J2351+2344 &  23 51 22.60 &  +23 44 19.37 &  -10.939302 \\
6772888139870271744 & 19312434-2134226 &               J1931-2134 &  19 31 24.40 & --21 34 24.43 &  -10.575118 \\
5314991556010305024 & 08412528-5736021 &               J0841-5736 &  08 41 25.25 &  --57 36 02.48 &  -10.126098 \\
3339914973377653760 & 05363846+1117487 &               J0536+1117 &  05 36 38.46 &  +11 17 47.84 &  -10.082494 \\
2430275225461072128 & 00274534-0806046 &               J0027-0806 &  00 27 45.47 &    --08 06 05.62 &   -9.931814 \\
4131758504393326592 & 16430128-1754274 &               J1643-1754 &  16 43 01.26 & --17 54 28.32 &   -9.657577 \\
4931155036049677312 & 01372781-4558261 & 2MASS J01372781-4558261 &   01 37 28.0 & --45 58 26.76 &   -9.046240 \\
5250988846733325312 & 09315840-6209258 &               J0931-6209 &  09 31 58.33 &  --62 09 25.46 &   -8.982967 \\
3240921406042422400 & 05200029+0613036 &                 THOR 11 &   05 20 00.31 &    +06 13 03.18 &   -8.062984 \\
6730618549223057664 & 18435838-3559096 &               J1843-3559 & 18 43 58.39 &  --35 59 10.10 &   -7.974694 \\
6668403989418902144 & 20320890-4742064 &               HD 195266 &  20 32 09.01 &  --47 42 08.64 &   -7.872895 \\
 527849041138749056 & 00270283+6630389 &               J0027+6630 &   00 27 03.02 &  +66 30 38.62 &   -7.723538 \\
 599277851962810112 & 08224748+0757171 &               J0822+0757 &  08 22 47.45 &   +07 57 16.26 &   -7.625252 \\
 628827128175694976 & 10141918+2104297 &                 MCC 124 & 10 14 19.03 &   +21 04 26.85 &   -7.252588 \\
 602637929434290304 & 08290411+1125053 &               J0829+1125 &   08 29 04.06 &   +11 25 04.31 &   -7.139063 \\
3240923360251087104 & 05203182+0616115 &                 THOR\ 10 &  05 20 31.83 &   +06 16 11.08 &   -7.116907 \\
4745195905754074496 & 02365171-5203036 &               J0236-5203 &  02 36 51.87 &   --52 03 03.61 &   -6.761954 \\
2854508592899528192 & 23500639+2659519 &               J2350+2659 &  23 50 06.62 &   +26 59 51.20 &   -6.458421 \\
 288290097672550528 & 01303534+2008393 &               J0130+2008 &  01 30 35.42 &   +20 08 38.82 &   -5.804100 \\
\hline
\end{tabular}
\end{center}
\end{table*}

\begin{table*}
Table~\ref{tab:rejects}, continued.\hfill\\
\vspace{1ex}
\noindent
\begin{center}
\begin{tabular}{|r|r|l|c|c|c|}
\hline
\multicolumn{1}{c}{\gaia\ DR3 ID} & \multicolumn{1}{c}{2MASS ID} & \multicolumn{1}{c}{Common Name} &  \multicolumn{1}{c}{RA (hh:mm:ss.ss)} &  \multicolumn{1}{c}{Dec (dd:mm:ss.s)} & \multicolumn{1}{c}{log P$_{BPMG}$}  \\
\hline
\hline
1915678735414682368 & 22571130+3639451 &               J2257+3639 & 22 57 11.38 &  +36 39 44.95 &   -5.576754 \\
3209947441933983744 & 05204041-0547109 &              BD-05 1229 &  05 20 40.42 &  -05 47 11.65 &   -4.692504 \\
  18211043587721088 & 02412589+0559181 &               BD+05 378 &  02 41 25.97 &   +05 59 17.52 &   -4.457175 \\
3002957677856537728 & 06322029-0943290 &        UCAC4 402-013866 &  06 32 20.28 &  --09 43 29.85 &   -3.610379 \\
2797571482766220160 & 00193931+1951050 & 2MASS J00193931+1951050 &   00 19 39.4 &   +19 51 04.37 &   -3.456861 \\
  59000206263788672 & 03105356+1838385 &               J0310+1838 &  03 10 53.63 &  +18 38 37.53 &   -3.308083 \\
4482593565100321664 & 18011345+0948379 &               J1801+0948 &  18 01 13.44 &   +09 48 37.36 &   -2.343479 \\
2962658549474035712 & 05064946-2135038 &           BD-21 1074 B &   05 06 49.48 &  --21 35 04.27 &   -2.284901 \\
 898486306258888832 & 07293108+3556003 &               J0729+3556 &  07 29 31.04 &   +35 55 58.60 &   -2.265637 \\
6104139604209342464 & 14252913-4113323 &         UCAC3 98-153320 & 14 25 29.06 & --41 13 33.05 &   -2.254027 \\
5226193446465971072 & 11091606-7352465 &               J1109-7352 &  11 09 15.86 & --73 52 46.38 &   -2.161118 \\
2797571448406493056 & 00194303+1951117 &               J0019+1951 &  00 19 43.12 &  +19 51 11.07 &   -2.156805 \\
4154524481067422976 & 18202275-1011131 &               J1820-1011 & 18 20 22.74 & --10 11 14.07 &   -2.074230 \\
3306841457454198016 & 04232720+1115174 &               J0423+1115 &  04 23 27.23 &  +11 15 16.84 &   -1.706733 \\
3623343812312995584 & 13215631-1052098 &               J1321-1052 & 13 21 56.25 & --10 52 10.79 &   -1.705694 \\
4946633857666189952 & 02303239-4342232 &               CD-44 753 &  02 30 32.42 & --43 42 23.39 &   -1.584435 \\
2556724418479276416 & 00464841+0715177 &               J0046+0715 &  00 46 48.52 &   +07 15 16.82 &   -1.575535 \\
3996306127913903232 & 10593834+2526155 &               BD+26 2161 & 10 59 38.11 &  +25 26 14.69 &   -0.896972 \\
3181961503752885248 & 05063003-1102350 &                HD 32965 &   05 06 30.08 &  --11 02 35.83 &   -0.701318 \\
 132592688469662208 & 02224082+3055161 &               J0222+3055 &  02 22 40.93 &   +30 55 15.0 &   -0.585495 \\
1905055700744817024 & 22184265+3321137 &               J2218+3321 & 22 18 42.73 &  +33 21 13.45 &   -0.472521 \\
6760846563417053056 & 18580415-2953045 &               J1858-2953 &  18 58 04.17 &  --29 53 05.42 &   -0.455103 \\
2640348977921037056 & 23314492-0244395 &                  AF Psc & 23 31 45.02 &  --02 44 40.66 &   -0.242801 \\
2507726817386336512 & 01561492+0006088 &        UCAC4 451-002420 &  01 56 15.04 &     +00 06 07.81 &   -0.157852 \\
2315841869173294080 & 00275023-3233060 &              GJ 2006 A &  00 27 50.37 &  --32 33 07.04 &   -0.147485 \\
6633474669669267712 & 19010683-5853301 &               HD 175897 &   19 01 6.86 & --58 53 30.88 &   -0.103548 \\
2962794236080240128 & 05082729-2101444 &               J0508-2101 &   05 08 27.33 &  --21 01 44.57 &   -0.094709 \\
 799092516692729344 & 09361593+3731456 &               J0936+3731 &  09 36 15.78 &  +37 31 44.12 &   -0.062335 \\
 \hline
\end{tabular}
\end{center}
\end{table*}

\bsp	
\label{lastpage}
\end{document}